\documentclass[preprint]{aastex}
\usepackage{color,natbib,lscape}
\newcommand{\HST}{{\sl HST}}

\newcommand{\Mjup}{\mbox{$M_{\rm Jup}$}}

\newcommand{\pertwo}{\mbox{$^{-2}$}}

\newcommand{\etal}{et al.}
\newcommand{\eg}{e.g.}
\newcommand{\ie}{i.e.}

\newcommand{\chisq}{\mbox{$\chi^2$}}

\newcommand{\htwoo}{{\hbox{H$_2$O}}}   % H20
\newcommand{\htwo}{{\hbox{H$_2$}}}     % H2
\newcommand{\meth}{{\hbox{CH$_4$}}}   % CH4
\newcommand{\ammonia}{{\hbox{NH$_3$}}}   % NH3

\newcommand{\degs}{\mbox{$^{\circ}$}}
\newcommand{\Lbol}{\mbox{$L_{\rm bol}$}}

\newcommand{\Teff}{\mbox{$T_{\rm eff}$}}
\newcommand{\logg}{\mbox{$\log(g)$}}

\newcommand{\cfbdsAB}{\hbox{CFBDSIR~J1458+1013AB}}

\newcommand{\cfbdsB}{\hbox{CFBDSIR~J1458+1013B}}
\newcommand{\cfbds}{\hbox{CFBDSIR~J1458+1013}}
\newcommand{\ugps}{\hbox{UGPS~J0722$-$0540}}

\newcommand{\WISE}{{\sl WISE}}

\newcommand{\eIndBab}{\hbox{$\epsilon$~Ind~Bab}}

\newcommand{\wiseone}{WISE~J1217+1626}
\newcommand{\wiseoneAB}{WISE~J1217+1626AB}
\newcommand{\wiseoneA}{WISE~J1217+1626A}
\newcommand{\wiseoneB}{WISE~J1217+1626B}

\newcommand{\wisetwo}{WISE~J1711+3500}
\newcommand{\wisetwoAB}{WISE~J1711+3500AB}
\newcommand{\wisetwoA}{WISE~J1711+3500A}
\newcommand{\wisetwoB}{WISE~J1711+3500B}

\slugcomment{ApJ, in press}
\shorttitle{Two Wide Binary Brown Dwarfs at the T/Y Transition}
\shortauthors{Liu et al.}

\begin{document}

\title{Two Extraordinary Substellar Binaries at the T/Y Transition and\\
  the $Y$-Band Fluxes of the Coolest Brown Dwarfs\altaffilmark{1,2}}


\author{Michael C. Liu,\altaffilmark{3,4}
Trent J. Dupuy,\altaffilmark{5,6}
Brendan P. Bowler,\altaffilmark{3}
S. K. Leggett,\altaffilmark{7}
William M. J. Best\altaffilmark{3}
}

\altaffiltext{1}{Most of the data presented herein were obtained at the
  W.M. Keck Observatory, which is operated as a scientific partnership
  among the California Institute of Technology, the University of
  California, and the National Aeronautics and Space Administration. The
  Observatory was made possible by the generous financial support of the
  W.M. Keck Foundation.} 
\altaffiltext{2}{Some of the observations were obtained at the Gemini
  Observatory, which is operated by the Association of Universities for
  Research in Astronomy, Inc., under a cooperative agreement with the
  NSF on behalf of the Gemini partnership: the National Science
  Foundation (United States), the Science and Technology Facilities
  Council (United Kingdom), the National Research Council (Canada),
  CONICYT (Chile), the Australian Research Council (Australia),
  Minist\'{e}rio da Ci\^{e}ncia, Tecnologia e Inova\c{c}\~{a}o (Brazil)
  and Ministerio de Ciencia, Tecnolog\'{i}a e Innovaci\'{o}n Productiva
  (Argentina).}
\altaffiltext{3}{Institute for Astronomy, University of
   Hawaii, 2680 Woodlawn Drive, Honolulu HI 96822}
\altaffiltext{4}{Visiting Astronomer at the Infrared Telescope Facility,
  which is operated by the University of Hawaii under Cooperative
  Agreement no. NNX-08AE38A with the National Aeronautics and Space
  Administration, Science Mission Directorate, Planetary Astronomy
  Program.}
\altaffiltext{5}{Harvard-Smithsonian Center for Astrophysics, 60 Garden
  Street, Cambridge, MA 02138}
\altaffiltext{6}{Hubble Fellow}
\altaffiltext{7}{Gemini Observatory, 670 North A'ohoku Place, Hilo, HI
  96720}

\begin{abstract} 
  \noindent Using Keck laser guide star adaptive optics imaging, we have
  found that the T9~dwarf \wiseone\ and T8~dwarf \wisetwo\ are
  exceptional binaries, with unusually wide separations
  ($\approx$0.8\arcsec, 8--15~AU), large near-IR flux ratios
  ($\approx$2--3~mags), and small mass ratios ($\approx$0.5)
  compared to previously known field ultracool binaries.
  Keck/NIRSPEC $H$-band spectra give a spectral type of Y0 for
  \wiseoneB, and photometric estimates suggest T9.5 for \wisetwoB.
  The \wiseoneAB\ system is very similar to the T9+Y0 binary \cfbdsAB;
  these two systems are the coldest known substellar multiples, having
  secondary components of $\approx$400~K and being planetary-mass
  binaries if their ages are $\lesssim$1~Gyr.
  Both \wiseoneB\ and \cfbdsB\ have strikingly blue $Y-J$ colors
  compared to previously known T~dwarfs, including their T9 primaries.
  Combining all available data, we find that $Y-J$ color drops
  precipitously between the very latest T~dwarfs and the Y~dwarfs. The
  fact that this is seen in (coeval, mono-metallicity) binaries
demonstrates that the color drop arises from a change in temperature,
not surface gravity or metallicity variations among the field
population.
  Thus, the T/Y transition established by near-IR spectra coincides with
  a significant change in the $\approx$1~\micron\ fluxes of ultracool
  photospheres.
  One explanation is the depletion of potassium, whose broad
  absorption wings dominate the far-red optical spectra of T~dwarfs.
  This large color change suggests that far-red data may be valuable
  for classifying objects of $\lesssim$500~K.
\end{abstract}

\keywords{binaries: general, close --- stars: brown dwarfs ---
  infrared: stars --- techniques: high angular resolution}

%----------------------------------------------------------------------%
\section{Introduction}

Binaries have played a central role in the study of the coolest dwarfs
over the last two decades. Each advance that has led to creation of a
later spectral type has benefitted from discovery of companions to
higher mass objects. The first L~dwarf, GD~165B, was found as a
companion to a white dwarf \citep{1988Natur.336..656B}. At the time, its
unusual optical spectrum compared to late-M dwarfs was puzzling, but
with the subsequent discovery of more L~dwarfs, it became apparent that
GD~165B was the prototype of a new spectral type
\citep{1999ApJ...519..834K}. The first T~dwarf, Gl~229B, was found
around an M~dwarf \citep{1995Natur.378..463N}, and its uniqueness was
evident from the strong \meth\ and \htwoo\ in its near-IR spectrum
\citep{1995Sci...270.1478O}.

Just last year, two low-mass companions were found with temperatures of
only $\approx$300--400~K, one being a tight (3~AU) companion to the
T9.5~dwarf \cfbds\ \citep{2011ApJ...740..108L} and the other being a
very wide (2500~AU) companion to the white dwarf GJ~3483
\citep{2011ApJ...730L...9L}. Spectroscopy of these objects has not been
possible yet, due to the small angular separation (0.11\arcsec) of
\cfbdsAB\ and the lack of a near-IR detection for GJ 3483B (a.k.a.
WD~0806$-$661B). However, their exceptionally faint absolute magnitudes
showed these two objects were novel compared to all previously known
T~dwarfs. Their temperatures, as inferred from their absolute magnitudes
using evolutionary models, are far cooler than the previous
record-holder, the $\approx$520~K T10~dwarf \ugps\
\citep{2010MNRAS.408L..56L}, and place the two companions in {\em terra
  incognita} where theoretical models predict the onset of new
photospheric signatures including \ammonia\ absorption in the near-IR,
the disappearance of neutral alkali lines, and the formation of water
clouds \citep{2003ApJ...596..587B}.
More recently, \citet{2011ApJ...743...50C} have identified five
free-floating objects with sufficiently distinct near-IR spectra to
propose classification as Y~dwarfs, based on the likely presence of
\ammonia\ absorption in the blue wing of their $H$-band continuua. While
reliable parallaxes are not yet available for these objects and thus
their absolute magnitudes are uncertain, model atmosphere fits indicate
temperatures of $\approx$300--500~K. Their novel near-IR appearance may
be a precursor to even more significant changes along the remaining
temperature gap down to Jupiter (124~K; \citealp{2005AREPS..33..493G}).

As part of our ongoing effort to study low-temperature objects using
substellar binaries, we present here the discovery of two
extraordinarily well-resolved binaries that shed light on the transition
from the latest T~dwarfs to the Y~dwarfs. The primaries,
WISEPC~J121756.91+162640.2 and WISEPA J171104.60+350036.8 (hereinafter
\wiseone\ and \wisetwo), were found from a mid-IR search of the solar
neighborhood by \citet{2011arXiv1108.4677K}, who asssigned
integrated-light near-IR spectral types of T9 and T8, respectively. We
also present improved near-IR imaging of \cfbdsAB. Our characterization
of these three binaries includes resolved photometry in the $Y$ band,
which spans the $\approx$1.0~\micron\ transmission window of the Earth's
atmosphere. A broadband filter in this window is a new capability for
adaptive optics imaging at Keck, just installed in September~2011. One
of the prime motivating factors for its procurement, as a collaboration
between one of us (M. Liu) and Keck Observatory, is to explore the
$\approx$1~\micron\ fluxes of the coolest substellar
  objects, which can be diagnostic of surface gravity, metallicity, and
photospheric chemistry \citep[e.g.][]{2012ApJ...748...74L}.

%----------------------------------------------------------------------%
\section{Observations \label{sec:observations}}

\subsection{Keck-II/NIRC2 Adaptive Optics Imaging}

We imaged \wiseone, \wisetwo, and \cfbds\ over 3~nights in January and
April~2012 using the laser guide star adaptive optics (LGS AO) system of
the 10-meter Keck~II Telescope on Mauna Kea, Hawaii
\citep{2006PASP..118..297W, 2006PASP..118..310V}. For \wiseone, we used
the facility IR camera NIRC2 with its wide field-of-view optics
($40.8\arcsec\times40.8\arcsec$), and for \wisetwo\ and \cfbds\ we used
the narrow field-of-view optics ($10.2\arcsec\times10.2\arcsec$).
Conditions were photometric, with 0.8\arcsec\ seeing in January and
0.4--0.6\arcsec\ seeing in April.
The LGS provided the wavefront reference source for AO correction, with
the tip-tilt motion measured simultaneously using
$R=15.5$, 14.4, and 15.5~mag stars from the USNO-B1.0 catalog
\citep{2003AJ....125..984M} located 38\arcsec, 43\arcsec, and 55\arcsec\
away from \wiseone, \wisetwo, and \cfbds, respectively.
The LGS brightness, as measured by the flux incident on the AO wavefront
sensor, was equivalent to a $V\approx 9.5-10.0$~mag star. The LGS was
pointed at the center of the NIRC2 field-of-view for all observations.

We obtained images with all or a subset of the broad-band
$Y$~(1.02~\micron), $J$~(1.25~\micron), $H$~(1.64~\micron), and
$K$~(2.20~\micron) filters of the Mauna Kea Observatories (MKO)
photometric system \citep{mkofilters1,mkofilters2}. We also obtained
data with the medium-band $CH4{s}$ filter (central wavelength of
1.592~\micron, 0.126~\micron\ wide), which is positioned around the
$H$-band peak in the spectra of T~dwarfs. Finally, for \wiseone, we
obtained data with NIRC2's custom ``$z$'' filter (central wavelength of
1.031~\micron, 0.048~\micron\ wide), which covers the cleanest portion
of the $Y$-band atmospheric window. To avoid confusion with optical
filters using the same letter but spanning different wavelengths, we
refer to this as the $z_{1.1}$ filter.

For each filter, we obtained a set of dithered images, offsetting the
telescope by a few arcseconds between every pair of images. The images
were analyzed in the same fashion as our previous work
\citep[e.g.][]{2006astro.ph..5037L, liu08-2m1534orbit}. The raw images
were reduced using standard methods for flat-fielding and
sky-subtraction. The binary's flux ratios and relative astrometry were
derived by fitting an analytic model of the point spread function as the
sum of three elliptical gaussians. 

For the very well-resolved \WISE\ binaries, we fitted all the individual
images and adopted the averages of the results as the final measurements
and the standard deviations as the errors. (The two \WISE\ binaries are
so wide that simple aperture photometry gave essentially the same
results, as the contamination of the secondary by the primary is
negligible.) For \cfbdsAB, we fitted the stacked mosaic to maximize the
S/N of the faint, close secondary and derived uncertainties from Monte
Carlo simulations, as we did in \citet{2011ApJ...740..108L}. For NIRC2's
wide camera, we adopted a pixel scale of $39.884\pm0.039$~mas/pixel and
an orientation for the detector's $+y$~axis of $0.16\pm0.09$\degs\
\citep{2006ApJ...649..389P}. For the narrow camera, we adopted a pixel
scale of $9.963 \pm 0.005$~mas/pixel and an orientation for the
detector's $+y$~axis of $0.13\pm0.07$\degs\
\citep{2008ApJ...689.1044G}.
The relative astrometry was corrected for (the very small) instrumental
optical distortion based on a solution by B. Cameron (priv. comm.).

\begin{figure}
\centerline{\includegraphics[width=8in,angle=0]{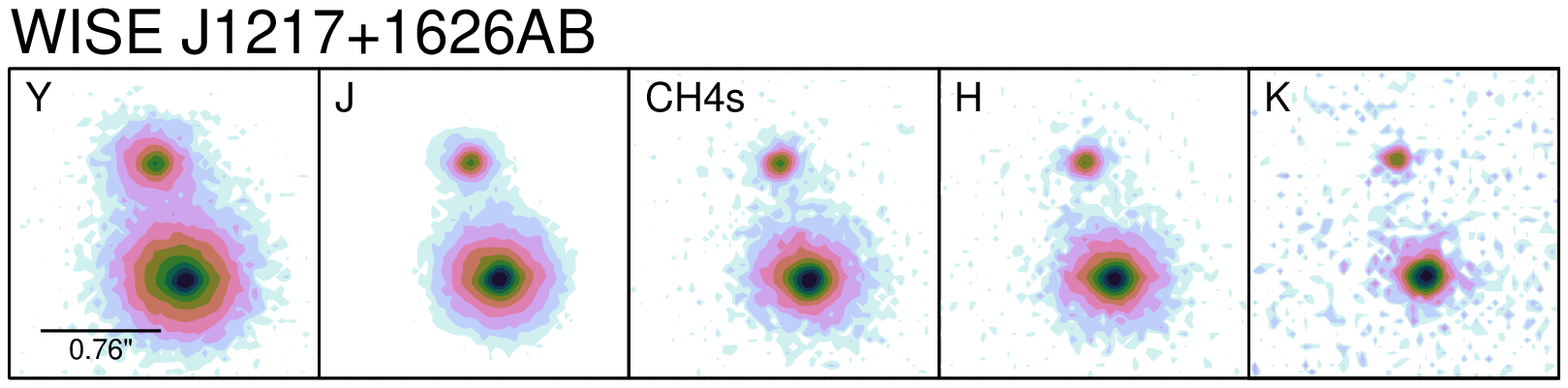}}
%\centerline{\includegraphics[width=8in,angle=0]{make-wise1217-fig.ps}}
\vskip 5ex
\centerline{\includegraphics[width=8in,angle=0]{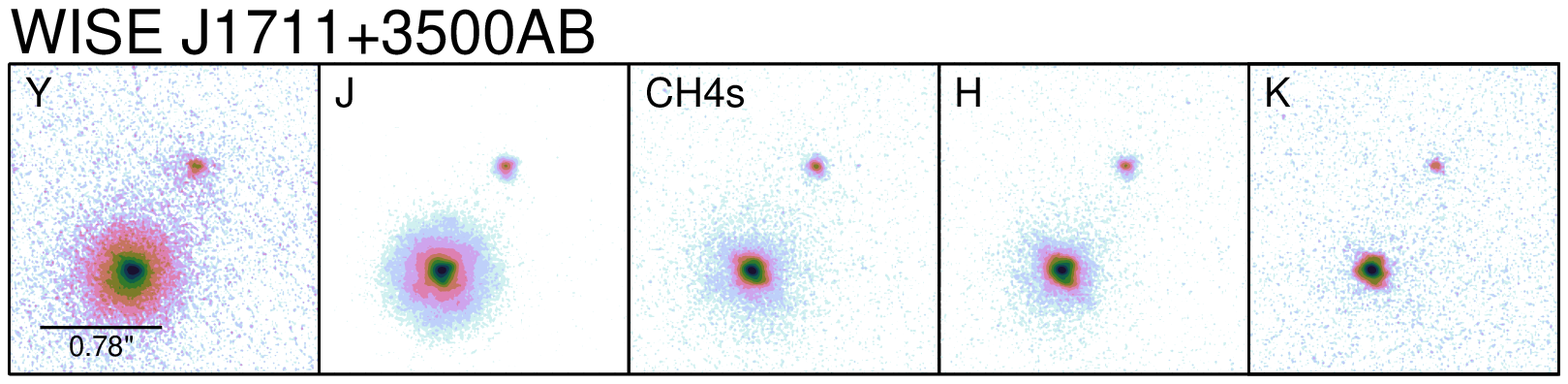}}
%\centerline{\includegraphics[width=8in,angle=0]{make-wise1711-fig.ps}}
\vskip 5ex
\centerline{\includegraphics[width=8in,angle=0]{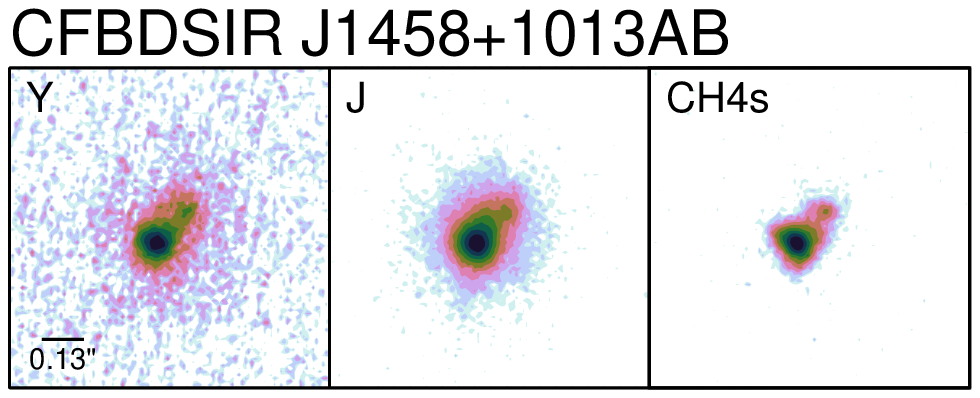}}
%\centerline{\includegraphics[width=8in,angle=0]{make-cfbds1458-fig.ps}}
%\vskip -2ex
\caption{\normalsize Keck LGS AO imaging of \wiseoneAB, \wisetwoAB, and
  \cfbdsAB. North is up and east is left. The images of the \WISE\
  binaries are 2.0\arcsec\ on a side with colored contours plotted in
  uniform logarithmic intervals in flux, from 100\% to 0.3\% of the peak
  flux in each bandpass. The image of \cfbdsAB\ is 1\arcsec\ on a side,
  with the lowest colored contour being 2\% of the peak (since the
  fainter magnitude of this object results in a lower S/N dataset). For
  \wiseoneAB, the $z_{1.1}$-band image is not shown, but it is
  essentially identical to the $Y$-band one. The PSF halo of
\wiseoneAB\ is better detected than for the other two binaries because
this dataset was obtained with NIRC2's most coarse pixel scale.
    \label{fig:images}}
\end{figure}

Table~\ref{table:keck} presents our final measurements, and
Figure~\ref{fig:images} shows our reduced images.
For all three binaries, the astrometry from the different filters shows
excellent agreement. For \wiseone, our measurements have \chisq\ values
of 0.89 and 3.46 (5 degrees of freedom) for the separation and PA,
respectively, when compared to the hypothesis of a constant value for
each quantity. For \wisetwo, the \chisq\ values are 0.2 and 3.10 (4
degrees of freedom) for separation and PA, respectively, and for
\cfbdsAB\ they are 2.3 and 0.17 (2 degrees of freedom).
Also, our \cfbdsAB\ $J$ and $CH4{s}$ flux ratios are consistent with
those obtained in 2010 by \citet{2011ApJ...740..108L}, with the new
$J$-band data being a significant improvement. Given that the
uncertainties in the resolved measurements are likely dominated by
systematic errors in deblending the two components of this tight binary,
we forgo averaging the 2010 and 2012 measurements and simply choose for
each filter the measurement with the smaller error in our analysis
(Section~3).

\subsection{Gemini/NIRI Photometry of \wiseone}

We obtained seeing-limited $Y$-~and $K$-band integrated-light photometry
of \wiseoneAB\ using the facility near-IR camera NIRI
\citep{2003PASP..115.1388H} on the Gemini-North 8.1-m telescope on Mauna
Kea, Hawaii. These observations complement the $J$- and $H$-band
photometry published in \citet{2011arXiv1108.4677K}. Our NIRI
observations were obtained as queue observing program GN-2012A-DD-2 on
01~April~2012~UT. The seeing was 0.7\arcsec\ FWHM, and the binary was
marginally resolved. We obtained five 60-s exposures at $Y$-band and
eighteen 30-s exposures at $K$-band, taken using a dither pattern with
10\arcsec\ offsets. Data were reduced in a standard fashion and
flatfielded with calibration lamps on the telescope. 
For photometric calibration, we observed the UKIRT faint standard FS~21,
using $K$-band magnitudes from \citet{2006MNRAS.tmp.1209L} and $Y$-band
magnitudes from the UKIRT online catalog.\footnote{\tt
  http://www.jach.hawaii.edu/UKIRT/astronomy/calib/phot$\_$cal/fs$\_$ZY$\_$MKO$\_$wfcam.dat}
We measured $Y_{NIRI}=18.55\pm0.03$~mag and $K=18.80\pm0.04$~mag for the
integrated-light system. As discussed in the Appendix, we apply a small
shift to transform the NIRI $Y$-band photometry to the MKO system,
giving a final value of $Y_{MKO} = 18.38\pm0.04$~mag.

% - - - - - - - - - - - - - - - - - - - - - - - - - -%
\subsection{Keck-II/NIRSPEC Spectroscopy of \wiseone}

We obtained resolved seeing-limited spectroscopy of \wiseoneAB\ with the
facilty near-IR spectrograph NIRSPEC \citep{1998SPIE.3354..566M} on the
Keck-II Telescope on 12~April~2012~UT. Seeing conditions were excellent,
with 0.5--0.7\arcsec\ FWHM in the optical as measured from the DIMM at
the nearby Canada-France-Hawaii Telescope, suitable to resolve the two
components of the binary. Very light cirrus was present in the sky. We
used the low spectral resolution mode with the
0.38\arcsec~$\times$~42\arcsec\ slit and the NIRSPEC-5 blocking filter,
which spans the $H$-band. The achieved spectral resolution
($\lambda/\delta\lambda$) was $2300\pm200$ based on measurements of
bright isolated sky emission lines. The slit was oriented along the
binary PA which did not match the parallactic angle. However, the binary
was observed at airmass 1.0--1.1 so systematic errors due to
differential chromatic refraction were negligible.

We obtained 12~exposures of \wiseoneAB, each with a~300~s integration
time, nodding the telescope in an ABBA fashion. Immediately afterwards,
we observed the A0V star HD~109055 for telluric calibration and then
obtained calibration observations of an internal flat-field source,
argon arc lamps, and a dark frame. During the spectroscopic observations
of both the science target and A0V calibrator, the NIRSPEC image quality
was afflicted by an unusual guiding program, causing the formation of
double-peaked PSFs in the spatial direction on the detector. The PSF's
secondary peak was typically about 2/3 the flux of the main peak and in
the direction opposite the binary's B~component.

Basic image reduction was performed with the REDSPEC reduction
package.\footnote{http://www2.keck.hawaii.edu/inst/nirspec/redspec}
This included bad pixel removal, spatial rectification of the images,
the creation of a wavelength solution, pairwise image subtraction, and
flat fielding. Residual sky line emission in the pair-subtracted images
was removed by subtracting a constant value at each wavelength as
measured from a spatial region 10 pixels in width on either side of the
target spectrum.

Because of the close separation of the binary pair and the unusual
double-peaked PSF, we extracted 1-d spectra from the 2-d rectified
images using custom IDL routines. We found that modeling the PSF as the
sum of two gaussians provided an excellent match to the data. The
2-gaussian PSF of the two binary components differed only in total flux
and position, as we fixed the FWHM and the relative amplitude of the
2~gaussians to be the same for each binary component.
To determine the PSF characteristics of each image, we fit this
2-gaussian PSF model to the collapsed 1.55-1.60~\micron\ spatial profile
of the individual reduced images using the nonlinear, least-squares
curve fitting package MPFIT written in IDL \citep{2009ASPC..411..251M}.
The fitting results were used to fix all parameters of the PSF except
the flux ratio of the binary. This model was then fit to the spatial
profile as a function of wavelength, while median-averaging the spectrum
in wavelength bins to boost the S/N. We found this model to be an
excellent fit to the data, as the subtraction of a 2-d image based on
the model fits from the original data produced essentially a pure noise
image, without any coherent residuals.

The resulting individual spectra for WISE~J1217+1626~A and~B were
computed at each wavelength bin by summing the flux of their respective
2-gaussian PSFs. To construct the final spectra, we combined the
individual extracted spectra by first scaling each one to the
median-averaged spectrum from 1.55--1.65~\micron. Then at each
wavelength we computed the weighted average, with the weights inversely
proportional to the FWHM of the PSF model. The purpose of the weighting
is to favor data obtained in better seeing conditions. We adopted the
standard error of the individual spectra as the measurement
uncertainties for the final spectrum.  

The spectra of the A0V calibrator star were extracted by summing flux in
a single fixed-sized aperture and then combined. This was used to
correct the spectra of \wiseoneA\ and~B for telluric absorption using
the {\tt xtellcor\_general} routine in the Spextool reduction package
for IRTF \citep{2004PASP..116..362C,2003PASP..115..389V}.
To test the sensitivity of the results to the fitting procedure, we
experimented with wavelength bin sizes of 5, 10, 15, and 20 pixels
(0.0014, 0.0028, 0.0042, 0.0055~\micron). We also extracted the spectra
by allowing the FWHM of each gaussian in the 2-gaussian PSF model vary.
The results were all consistent within the spectral measurement
uncertainties. We use the 10-pixel binned data for our final analysis,
as the resulting sampling gives a resolution ($R\approx200$) compared to
other published near-IR spectra of late-T and Y~dwarfs. 

\begin{figure}
\vskip -1in
\centerline{\includegraphics[width=4.5in,angle=90]{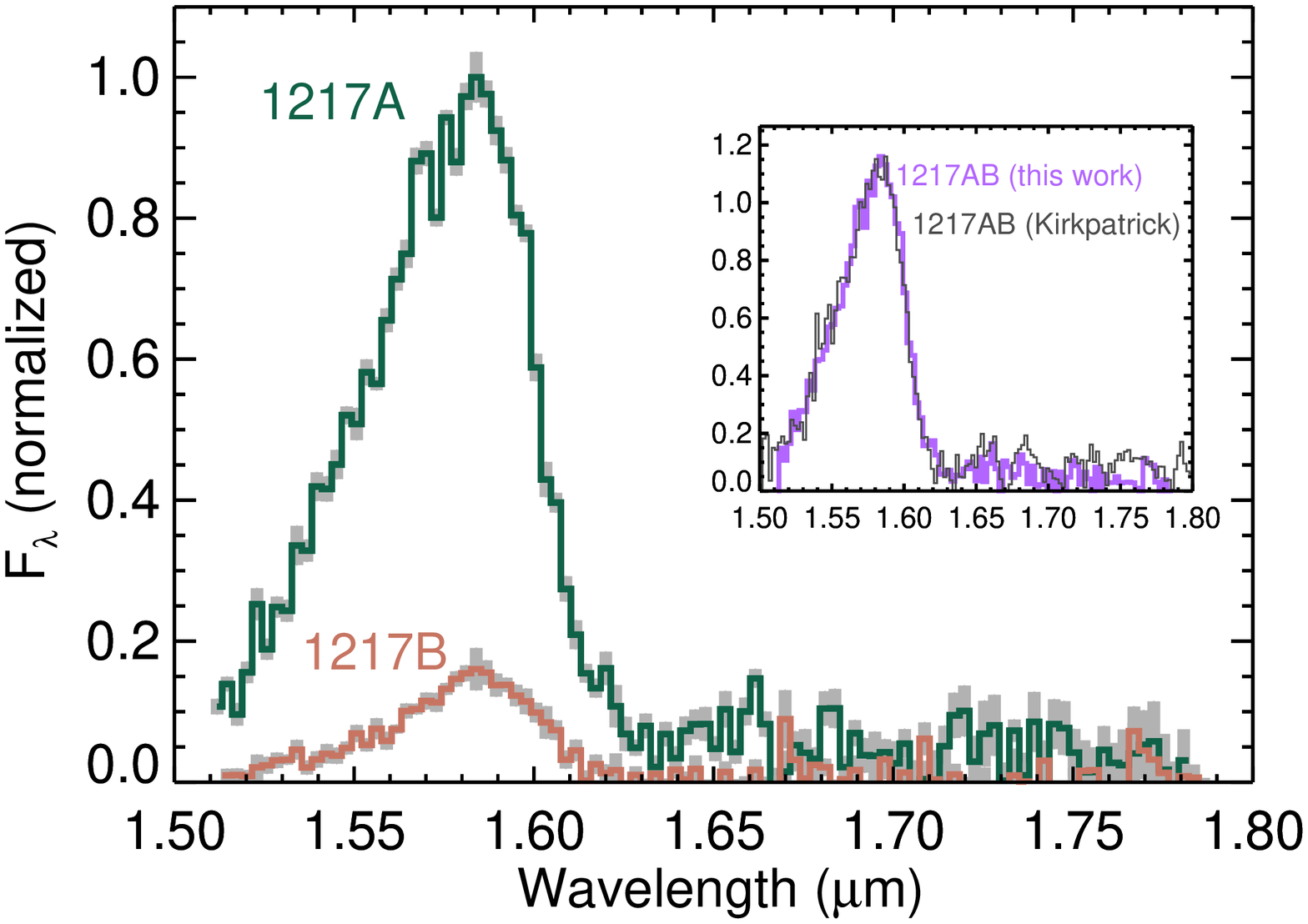}}
%\centerline{\includegraphics[width=4.5in,angle=90]{plot-spectra-wise1217.ps}}
\vskip 2ex
\caption{\normalsize Keck/NIRSPEC $H$-band spectra of the two components
  of \wiseoneAB, normalized to the peak flux of component~A. The grey
  shaded histograms enswathing the spectra give the measurement
  uncertainties. The inset figure shows the sum of our NIRSPEC spectra
  for the components in purple, with the integrated-light spectrum from
  \citet{2011arXiv1108.4677K} shown in grey. The latter has been median
  smoothed by an 11-pixel box to boost its S/N and rebinned to
  comparable sampling as our NIRSPEC data. \label{fig:spectra}}
\end{figure}

Figure~\ref{fig:spectra} shows the reduced spectra of the two
components. The peak S/N are about 60 and 15 for the primary and
secondary, respectively. As a cross-check on our reduction, we also
compare the sum of our spectra with the integrated-light spectrum from
\citet{2011arXiv1108.4677K} and find excellent agreement, as shown in
the inset of Figure~\ref{fig:spectra}. We also compute an $H$-band flux
ratio from our NIRSPEC spectra of $2.25\pm0.08$~mag, in accord with our
Keck LGS imaging ($2.20\pm0.04$~mag).

% - - - - - - - - - - - - - - - - - - - - - - - - - -%
\subsection{IRTF/SpeX Photometry and Spectroscopy of \wisetwo \label{sec:spex}}

To improve upon the modest S/N data for \wisetwo\ presented in
\citet{2011arXiv1108.4677K}, we obtained near-IR photometry and
low-resolution ($R\approx100$) spectroscopy of \wisetwo. Such data are
important for our analysis, since the Keck LGS imaging provides only
flux ratios for the two components. Computing the resolved magnitudes
requires multi-band integrated-light photometry, which can be
synthesized from a near-IR spectrum and photometry in a single band.

We used the facilty instrument SpeX \citep{1998SPIE.3354..468R} at
NASA's Infrared Telescope Facility located on Mauna Kea, Hawaii on
2012~April~20~UT. For imaging, we obtained 9~dithered exposures of
\wisetwo\ and the UKIRT faint standard FS~27 using the MKO $J$-band
filter. Conditions were photometric with good seeing, 0.6--0.7\arcsec\
FWHM in $J$-band as measured from the reduced images. Data were reduced
in a standard fashion, and we used aperture photometry
to determine an apparent magnitude of $J_{MKO} = 17.59\pm0.03$~mag,
with the uncertainty determined from the quadrature sum of the standard
error of the fluxes from the individual images of the science target and
photometric calibrator. \citet{2011arXiv1108.4677K} reported $J_{2MASS}
= 17.89\pm0.13$~mag, which is consistent with our measurement given the
$\approx$0.3~mag offset expected for late-T dwarfs between the MKO and
2MASS filter systems \citep{2004PASP..116....9S}. Our near-IR spectrum
(described below) finds $J_{2MASS} - J_{MKO} = 0.26$~mag, consistent
with the $0.30\pm0.13$~mag difference between our photometry and that of
Kirkpatrick \etal

\begin{figure}
\vskip -1in
\centerline{\includegraphics[width=5in,angle=90]{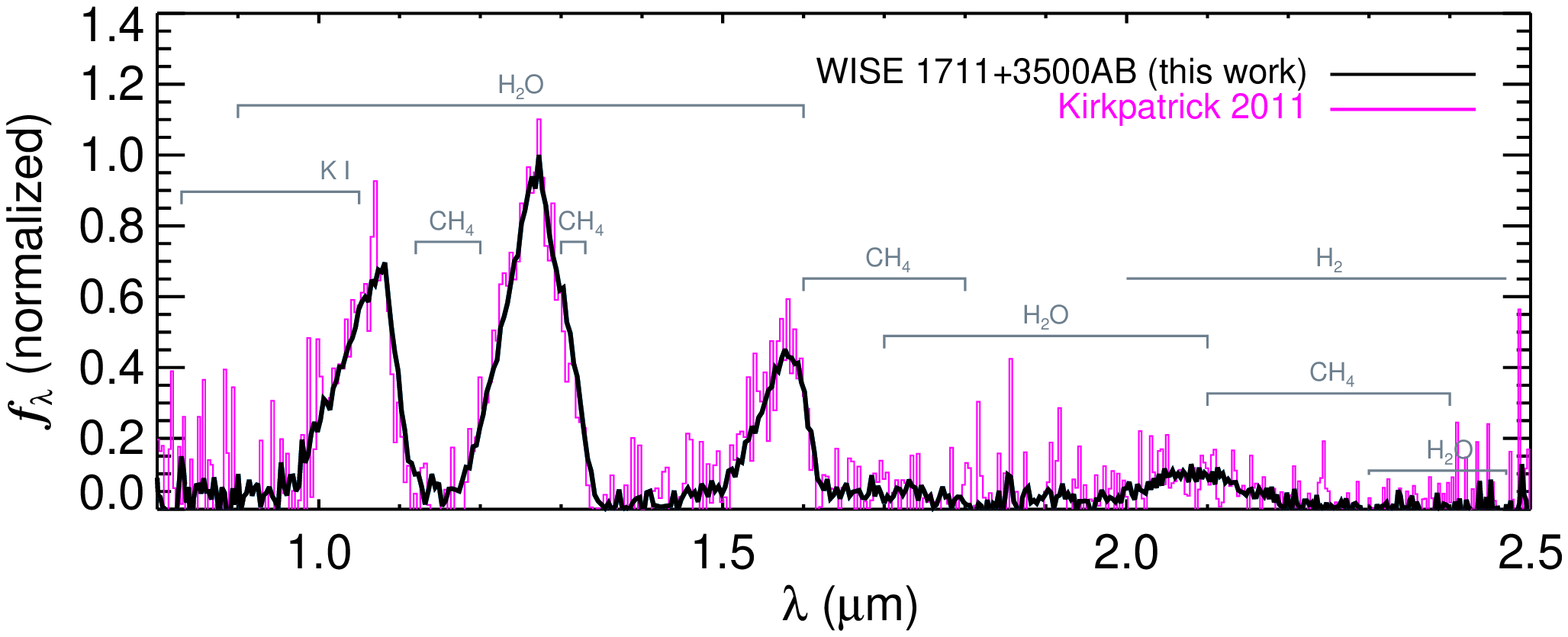}}
%\centerline{\includegraphics[width=5in,angle=90]{plot-spectra-wise1711.ps}}
\vskip -1in
\caption{\normalsize Our IRTF/SpeX integrated-light prism spectrum
  of \wisetwoAB\ compared with that of \citet{2011arXiv1108.4677K}.
  Major molecular features are marked in grey.  The spectra are
  normalized to unity flux at the peak of their $J$-band continuua. 
  \label{fig:spectra1711}}
\end{figure}

For spectroscopy, we used Spex in prism mode with the 0.8\arcsec\ slit,
obtaining 0.8--2.5~\micron\ spectra in a single order. 
\wisetwo\ was nodded along the slit in an ABBA pattern with individual
exposure times of 180~sec for a total exposure time of 48~min and
observed over an airmass range of 1.04--1.06.
We observed the A0V star GAT~7 \citep{1973AJ.....78..769G}
contemporaneously for telluric calibration. All spectra were reduced
using version~3.4 of the SpeXtool software package
\citep{2003PASP..115..389V,2004PASP..116..362C}. The S/N per pixel in
the final reduced spectrum is about 35, 50, 25, and 8 in the $YJHK$
peaks, respectively.
Figure~\ref{fig:spectra1711} shows that our spectrum agrees well with
that of \citet{2011arXiv1108.4677K}. We use our spectrum and $J$-band
photometry to synthesize the integrated-light magnitudes in the
$YHK$~bands on the MKO system (Table~\ref{table:wisetwo}).

%----------------------------------------------------------------------%
\section{Results \label{sec:results}}

\subsection{Evidence for Companionship}

With only one epoch of imaging, it is not possible to determine if the
companions seen in the Keck LGS images are co-moving with the our two
\WISE\ targets. But the resolved photometry of the systems (and spectra
in the case of \wiseoneAB) provides overwhelming circumstantial evidence
that the binaries are true physical associations. Specifically, the
nearly identical flux ratios in the $CH_4s$ and $H$-band images
indicates that the secondary components also have strong methane
absorption, with basically no flux redward of the 1.6~\micron\ flux
decrement seen in T~dwarfs. The possibility of two unassociated T~dwarfs
having such a small angular separation is diminishingly small.
\citet{2008MNRAS.391..320B, 2010MNRAS.406.1885B} identified 54~T dwarfs
in an area of 700~degs$^2$ to a depth of $J=19.0$~mag. Nine of these
objects are spectral type T7 or later, where the $H$-band methane
absorption is saturated in medium-band filter photometry (Figure~4 of
\citealp{liu08-2m1534orbit}). Assuming a uniform space density of
late-T~dwarfs, this means a surface density of 0.20~late-T~dwarfs
per~degs\pertwo\ to a depth of $J=21.0$~mag. Within the field of view of
the NIRC2 wide and narrow cameras, this corresponds to probabilities of
$2.6\times10^{-5}$ and $1.6\times10^{-6}$ for an unassociated T~dwarf to
be next to \wiseone\ and \wisetwo, respectively. This is highly
unlikely.

Note that this calculation is an {\em a posteriori} one based on a
discovery of a companion to a individual target. In fact, we need to
consider the total number of objects that we have imaged with
sensitivity to late-T dwarf companions. Our Keck LGS program to date
\citep[e.g.][]{2006astro.ph..5037L, 2010liu-2m1209, 2011ApJ...740..108L}
has targeted about 100~field T~dwarfs using the NIRC2 narrow camera.
Thus the probability of an unassociated late-T companion being
contiguous to any single object in our survey is $1.6\times10^{-4}$,
respectively. However, note that both these ``single-object'' and
``total-survey'' calculations overstate the odds of a chance
association, since they do not consider that the photometric distance of
a background late-T dwarf would not be compatible with those of the
science targets. We conclude that both \WISE\ systems are physical
binaries.

We searched for common proper motion companions within 10$\arcmin$ of
\wiseone, using the proper motion reported in Kirkpatrick et al. (2011).
We found no comoving objects in the Hipparcos, Tycho, LSPM-N, or NLTT
catalogs. The reported proper motion of \wisetwo\ is consistent with
zero, obviating a search for comoving companions.

% - - - - - - - - - - - - - - - - - - - - - - - - - - - - - - - -%
\subsection{Photometric Distances \label{sec:distance}}

We can estimate photometric distances to our two binaries using the
late-T primary components, and then as a result we also know the
distances to the T/Y secondary components. To compute a photometric
distance for \wiseoneAB, we compare the apparent magnitudes of
component~A to the six T8.5 and T9~dwarfs with parallaxes summarized in
\citet{2012arXiv1201.2465D} and using the \citet{2011ApJ...743...50C}
spectral types for the latest T~dwarfs. We include the T8.5~dwarfs
(1)~to compensate for the small sample of only two T9~dwarfs with
parallaxes and MKO near-IR photometry (UGPS~J0722$-$0540 and
CFBDSIR~J1458+1013A, which differ from each other by $\approx$0.9~mag in
absolute magnitudes) and (2)~to encompass the current uncertainties in
the spectral typing of the very coolest brown dwarfs. A weighted average
of the absolute magnitudes gives $M(Y, J, H, K) = 18.74\pm0.38,
17.88\pm0.35, 18.21\pm0.34, 18.42\pm0.39$~mag, where the uncertainty
here is the RMS scatter of the six objects. We use the $J$- and $H$-band
data for the photometric distance, as the $Y$~and $K$~bands are more
influenced by variations in metallicity and surface gravity
\citep[e.g.][]{2007ApJ...667..537L, 2006liu-hd3651b}. Both bands give an
identical distance modulus of $(m-M) = 0.10\pm0.35$~mag
($10.5\pm1.7$~pc) for \wiseoneA.
  
In a similar fashion, for \wisetwoAB\ we use the
\citet{2012arXiv1201.2465D} weighted average of $J$- and $H$-band
absolute magnitudes for T8~dwarfs, choosing the three objects that are
``normal'' (not young or low metallicity). These give distance moduli of
$1.35\pm0.35$ and $1.41\pm0.32$~mag, respectively. We take the average
and adopt the slightly larger RMS for a final distance modulus of
$1.38\pm0.35$~mag ($19\pm3$~pc) for \wisetwoA.

As expected, our photometric distances are larger than the 6.7~pc and
17.0~pc estimates for \wiseone\ and \wisetwo, respectively, given by
\citet{2011arXiv1108.4677K} which were based on integrated-light
$W2$-band photometry assuming a single object. Note that if the large
flux ratios we observe in the near-IR ($\approx$2--3~mag) were also
representative of the mid-IR flux ratios, we would expect better
agreement between our distances and those of Kirkpatrick \etal\ In other
words, the secondary components in these systems must be relatively
brighter in the mid-IR and thus contribute a larger portion of the
$W2$-band flux, as expected based on the very red mid-IR colors of the
coolest brown dwarfs \citep[e.g.][]{2010ApJ...710.1627L}.

% - - - - - - - - - - - - - - - - - - - - - - - - - - - - - - - -%
\subsection{Spectral Types \label{sec:spectra}}

\subsubsection{\wiseoneAB}

\begin{figure}
%\vskip -1in
\hbox{
\hskip -0.3in
\includegraphics[width=3in,angle=0]{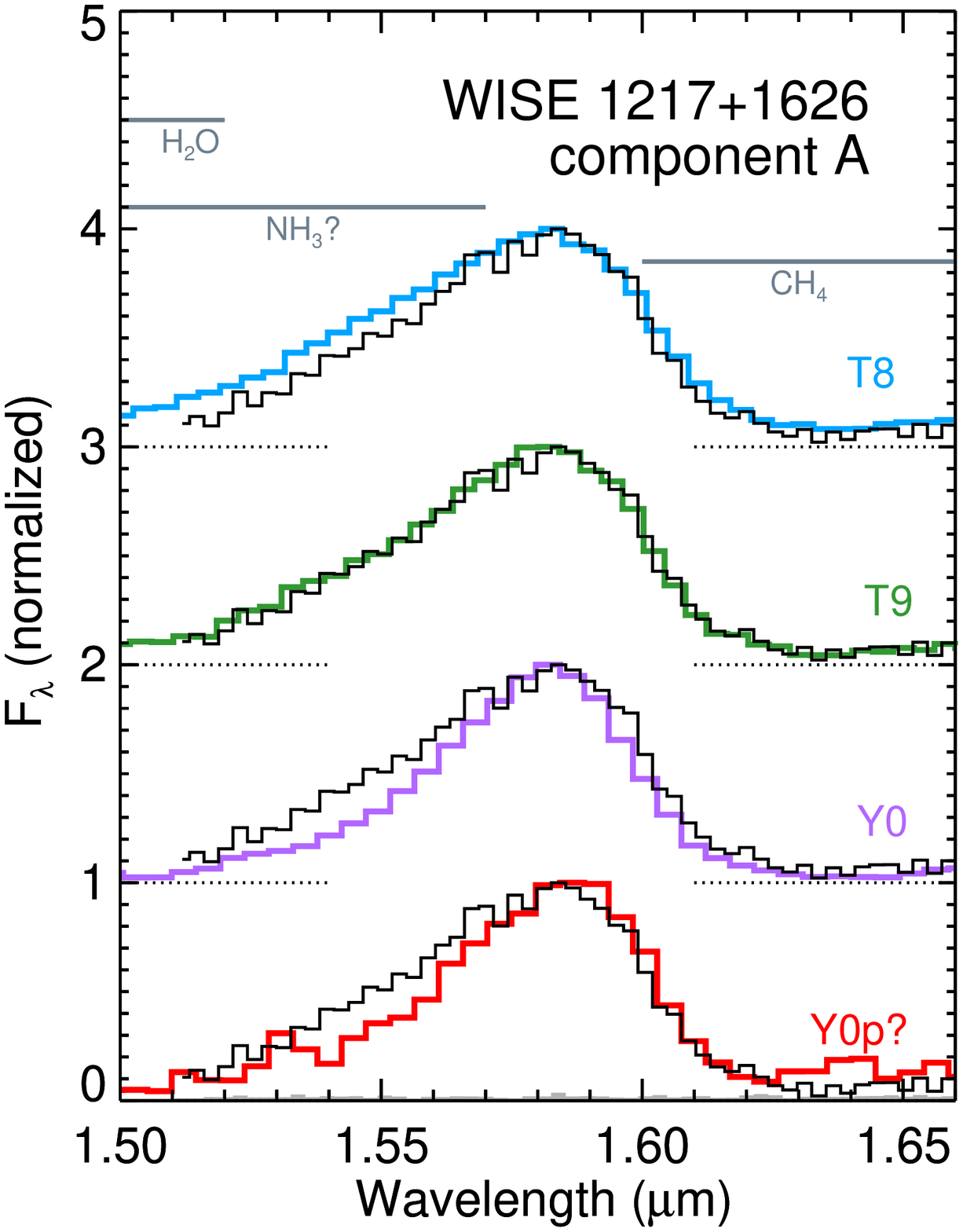}
\hskip 0.5in
\includegraphics[width=3in,angle=0]{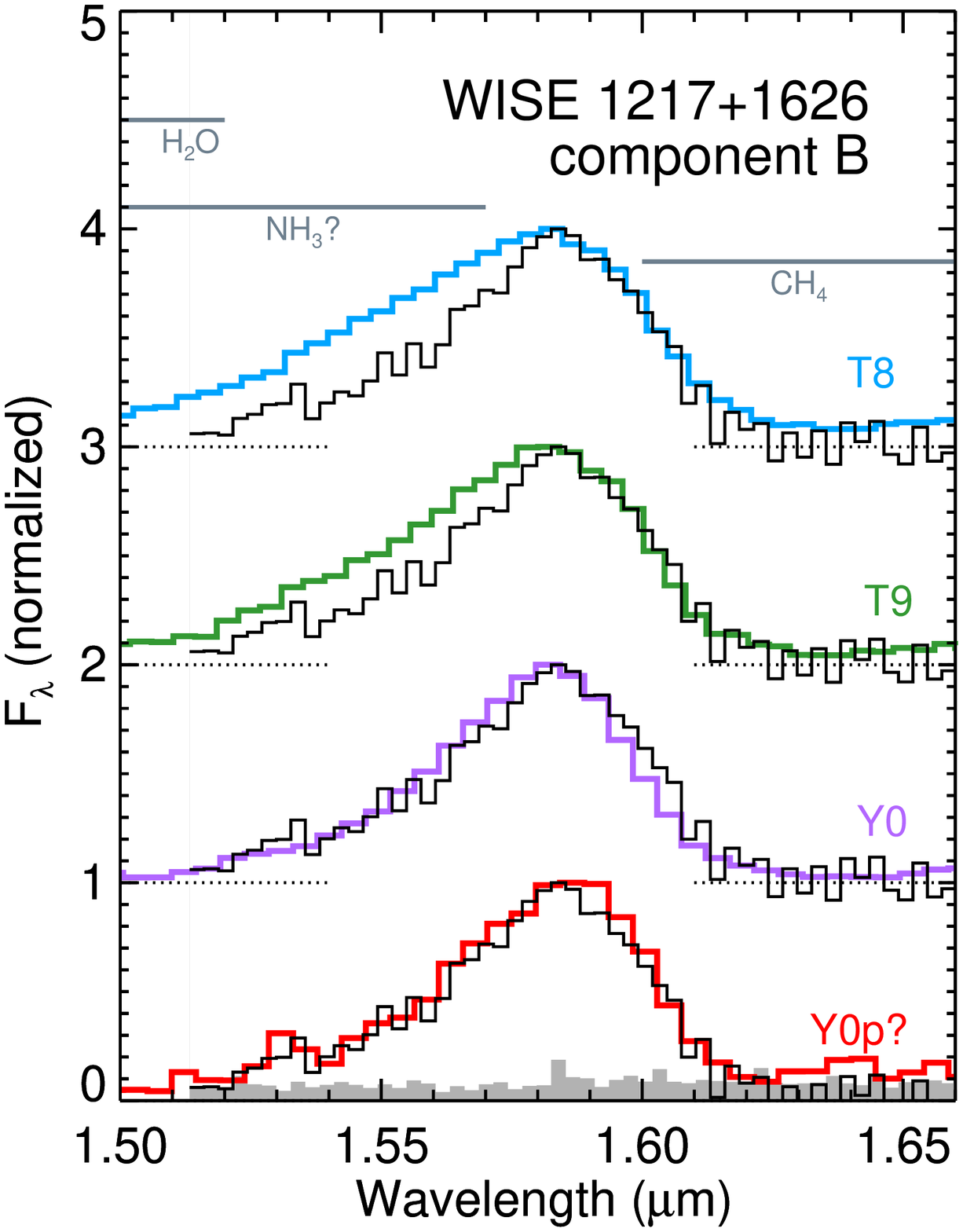}
%\includegraphics[width=3in,angle=0]{plot-typing-wise1217-alt-A.ps}
%\hskip 0.5in
%\includegraphics[width=3in,angle=0]{plot-typing-wise1217-alt-B.ps}
}
\vskip 2ex
%\vskip -1in
%\centerline{\includegraphics[width=5in,angle=90]{plot-typing-wise1217.ps}}
%\vskip 2ex
\caption{\normalsize  $H$-band spectra of the two components
  of \wiseoneAB\ compared to spectral standards 2MASS~J0415$-$0935 (T8),
  \ugps\ (T9), and WISE~J1738+2732 (Y0) as well as the peculiar Y0~WISE~1405+5534.
  All the spectra have been normalized to their peak flux.  The grey
  shaded histogram at the 
  bottom of each panel shows the measurement uncertainties for the
  \wiseoneAB\ spectra. 
  \label{fig:typing}}
\end{figure}

Our NIRSPEC spectra of both components of \wiseoneAB\ exhibit the deep
\htwoo\ and \meth\ absorption in the $H$-band that is characteristic of
the coolest brown dwarfs. Figure~\ref{fig:typing} compares our $H$-band
spectra with the T9 and Y0 spectroscopic standards proposed by
\citet{2011ApJ...743...50C}. Component~A shows excellent agreement with
their T9 standard \ugps. The spectral type for this component is
identical to the integrated-light type, as expected given the large
near-IR flux ratios.

Component~B shows good match with the Y0~standard WISE~J1738+2732. In
particular, the hallmark of the Y~spectral type proposed by
\citet{2011ApJ...743...50C} is the enhanced absorption at
1.53--1.58~\micron\ seen in the low-resolution spectra of the coolest
\WISE\ discoveries that is distinct from the latest T~dwarfs. They
attribute this feature to \ammonia, though a definitive identification
is still pending.
As seen in Figure~\ref{fig:typing}, \wiseoneB\ shows the same enhancd
absorption on the blue side as the Y0~standard, and thus we classify it
as Y0.

The red side of the $H$-band continuum for \wiseoneB\ does not appear to
be as sharply truncated as for the Y0~standard. However, the other
Y0~dwarf from the \citet{2011ApJ...743...50C} with reasonable S/N,
WISE~J1405+5534, also has a similar redward extent, which motivated a
classification of ``Y0~(pec?)'' in their discovery paper. Given the
modest S/N of our spectrum, we defer a decision on whether \wiseoneB\ is
spectroscopically peculiar.
Followup resolved spectroscopy with broader wavelength coverage will
help refine the spectral typing, especially as the width of the $J$-band
continuum appears to decrease with later-type objects.

\subsubsection{\wisetwoAB}

In the absence of resolved spectroscopy for \wisetwoAB, we consider the
flux ratios and estimated absolute magnitudes of the components to
estimate their spectral types. The large near-IR flux ratios
($\approx$2.8~mag) suggests strongly that the T8 integrated-light
spectrum is dominated by the primary. This is corroborated with simple
numerical experiments combining spectra of late-T and Y0~dwarfs using
the observed $H$-band flux ratio as a constraint. The $>$10$\times$
fainter secondary has negligible impact on the near-IR spectrum. Thus we
safely classify \wisetwoA\ as T8.

For the secondary, we use its near-IR absolute magnitude as estimated
from the photometric distance to the primary. This is obviously an
uncertain process, though preliminary parallaxes from
\citet{2012arXiv1205.2122K} suggest that the near-IR absolute magnitudes
drop off quickly at the T/Y transition, which lessens the impact of
uncertainties in the photometric distance. \wisetwoB\ has
$M(H)=19.58\pm0.36$~mag, which is intermediate between the two T9 and
two Y0~dwarfs with parallaxes in Figure~12 of
\citet{2012arXiv1205.2122K}. (Their figure has \cfbdsB\ plotted as the
sole T9.5, but as discussed in the next subsection this object is more
likely to be typed as Y0.) In terms of its $Y-J$ color and estimated
color-magnitude diagram position, \wisetwoB\ also appears to be
intermediate between the T9 and Y0 dwarfs (Section~\ref{sec:phot}). Thus
we estimate a spectral type of T9.5.

\subsubsection{\cfbdsAB}

At the time \cfbdsAB\ was discovered to be a binary, the near-IR
absolute magnitudes of its secondary component were fainter than the
coolest known dwarf, \ugps\ which was classified as T10 at the time
\citep{2010MNRAS.408L..56L}. This motivated an initial spectral type
estimate of $>$T10 for the secondary by \citet{2011ApJ...740..108L},
with the primary type assumed to be identical to the integrated-light
type of T9.5. While the secondary type was unknown given the absence of
resolved spectroscopy, the monotonic behavior of near-IR absolute
magnitudes among the T~dwarfs demonstrated the object was exceptionally
low temperature. Since then, there have been several late-T and Y~dwarfs
identified from the \WISE\ dataset, which has warranted a
reclassification of the integrated-light spectrum to T9 by
\citet{2011ApJ...743...50C}. Given this new state of knowledge, we
reexamine here the resolved spectral types..

Based on the latest $H$-band absolute magnitudes from
\citet{2012arXiv1201.2465D}, \citet{2011arXiv1108.4677K} estimated types
of T8.5 and T9.5 for the two components of \cfbdsAB. However, closer
examination of the data indicates these types are too early. As is the
case with \wisetwoA, the large near-IR flux ratios ($\approx$2~mag) of
\cfbdsAB\ indicate the T9 integrated-light spectrum is dominated by the
primary. This is corroborated by the aforementioned experiments of
summing spectra and by the model atmosphere fitting in
\citet{2011ApJ...740..108L}, which found a blended-light spectrum had
negligible impact on the derived physical parameters. As for the
secondary, it has $M(H) = 19.99\pm0.23$ mag, in good agreement with the
two Y0~dwarfs in \citet{2012arXiv1205.2122K} with parallaxes. This
absolute magnitude also agrees well with that estimated for the
(spectroscopically typed) Y0 dwarf \wiseoneB, based on our photometric
distance for \wiseoneA. Thus we estimate the near-IR spectral types for
the two components of \cfbdsAB\ are T9 and Y0.

% - - - - - - - - - - - - - - - - - - - - - - - - - - - - %
\subsection{Resolved Photometry\label{sec:phot}}

\begin{figure}
\vskip -1in
\centerline{\includegraphics[width=6in,angle=90]{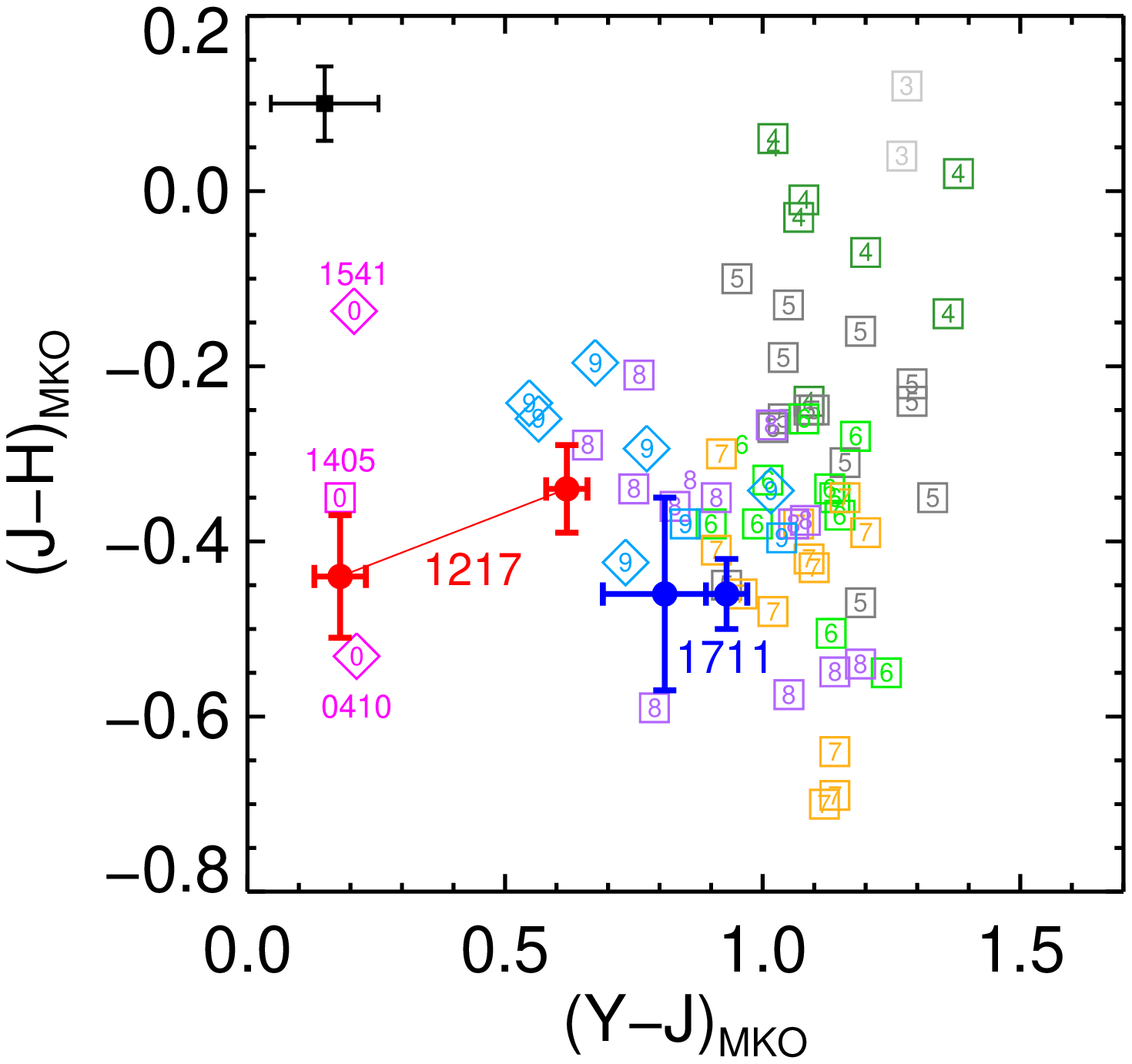}}
%\centerline{\includegraphics[width=6in,angle=90]{plot-colors.ps}}
\vskip -2ex
\caption{\normalsize Resolved near-IR colors of \wiseoneAB\ and
  \wisetwoAB\ compared with published photometry of field T~dwarfs
  (squares) and our synthetic photometry (diamonds) for T9--Y0~dwarfs.
  (See Section~\ref{sec:phot} for references.) For our \WISE\ binaries,
  the secondary components for both systems are the points plotted to
  the left, being bluer in $Y-J$ than the primaries. (The resolved
  colors of \cfbdsAB\ are consistent with the two \WISE\ binaries, but
  are not plotted given their much larger uncertainties.) The plotted
  numbers indicate the near-IR subclass of the objects, with half
  subclasses being rounded down (\eg, ``3'' represents T3 and T3.5
  dwarfs and ``0'' are the two Y0 dwarfs). Objects of the same subclass
  are also plotted in the same color. Known binaries are plotted as bare
  numbers without a encompassing square. The symbol in the upper left
  shows the median photometry errors for the field
  sample.\label{fig:colorcolor}}
\end{figure}

Tables~\ref{table:wiseone},~\ref{table:wisetwo}, and~\ref{table:cfbds}
summarize the resolved photometry for our binaries, derived from the
Keck LGS imaging and the available $YJHK$ integrated-light photometry.
Figure~\ref{fig:colorcolor} shows the $YJH$ color-color plot comparing
our binary sample to the known field T~dwarfs. For the field sample, we
use the data for 80~T~dwarfs compiled by \citet{2010ApJ...710.1627L},
one late-T~dwarf (WISE~J1617+1807) from
\citet{2011arXiv1108.4677K},\footnote{The T9~dwarf WISE~J1614+1739 in
  Kirkpatrick \etal\ also has $YJH$ photometry but the resulting colors
  are unusual, with $Y-J=0.35\pm0.12$ and $J-H=0.61\pm0.22$~mag, which
  place it far from the color locus of all other T~dwarfs. (The object
  would lie off the boundaries of Figure~\ref{fig:colorcolor} to the
  upper left.). However, the near-IR spectrum appears similar to other
  late-T dwarfs, and synthetic photometry of the spectrum gives
  $Y-J=0.77$~mag and $J-H=-0.26$~mag. Thus the photometry seems
  erroneous, so we exclude it.} three recent late-T discoveries from the
UKIDSS survey \citep{burningham08-T8.5-benchmark, 2010MNRAS.408L..56L,
  2011MNRAS.414.3590B}, and one Y0~dwarf (WISE~J1405+5534) from
\citet{morley12-clouds}.\footnote{The Morley \etal\ $Y$-band photometry
  was obtained with Gemini/NIRI, so we apply the shift computed in the
  Appendix to transform it to the MKO system.} For the objects
originally classified as T8.5--T10 in the Leggett \etal\ compilation, we
use the slightly earlier spectral types of T8--T9 proposed by
\citet{2011ApJ...743...50C}.
We also queried the UKIDSS\footnote{UKIDSS (UKIRT Infrared Deep Sky
  Survey) is described in \citet{2007MNRAS.379.1599L}. UKIDSS uses the
  UKIRT Wide Field Camera (WFCAM; \citealp{2007A&A...467..777C}), and a
  photometric system described in \citet{2006MNRAS.367..454H}. The
  pipeline processing and science archive are described in Dye \etal\
  (2006) and \citet{2008MNRAS.384..637H}.} Large Area Survey Data
Release~9 (DR9, Table~\ref{table:ukidss}) in order to add photometry for
five late-T dwarfs from Kirkpatrick \etal\ and for the T8~dwarf
PSO~J043.5+02 (a.k.a. WISE~J0254+0223, found independently by
\citealp{2011A&A...532L...5S} and \citealp{2011ApJ...740L..32L}).
Finally, we computed $YJH$ synthetic photometry for the six T9--T9.5 and
two Y0~dwarfs (WISE~J0410+1502 [Y0] and WISE~J1541$-$2250 [Y0.5]) in
\citet{2011arXiv1108.4677K} and \citet{2011ApJ...743...50C} with
sufficently blue coverage in their spectra to include the $Y$-band.

While \wiseoneA, \wisetwoA, and \wisetwoB\ reside in the loci of the
coolest known objects, the unusual nature of \wiseoneB\ is apparent in
the $YJH$ color-color diagram, being distinguished by its very blue
$Y-J$ color compared to all previously known T~dwarfs. Only the three
field Y~dwarfs in our sample are comparable. This blueness is also
emphasized in the near-IR color-magnitude diagram
(Figure~\ref{fig:cmd}), where component~B is fainter and significantly
bluer in $Y-J$ than any T~dwarf, based on the photometric distance for
the binary. The resolved $Y-J$ colors for \cfbdsAB\ show a similarly
pronounced blue difference between the primary and secondary component.
This is seen most clearly in the relative flux ratios, $\Delta(Y-J) =
0.43\pm0.02$~mag for \wiseoneAB\ and $0.47\pm0.15$~mag for \cfbdsAB.
(The integrated-light $Y$-band photometry for \cfbdsAB\ from UKIDSS is
only $S/N=5$ so comparing the resolved colors of the two components is
more noisy than comparing their flux ratios, which depend only on the
Keck LGS data.)

To further highlight this phenomenon, Figure~\ref{fig:yj} shows the
$Y-J$ colors of the coolest brown dwarfs as a function of spectral type,
based on published photometry and our synthesized photometry (regardless
of whether the objects have parallaxes or not).
\citet{2010ApJ...710.1627L} and \citet{2010MNRAS.406.1885B} have noted
that the $Y-J$ colors of the latest~T dwarfs (T8--T9) are bluer than the
earlier-type objects, and \citet{2011ApJ...743...50C} found that the
$Y$-band peak of two Y0~dwarfs are relatively brighter compared to the
T9~standard \ugps. Figure~\ref{fig:yj} shows that in fact the blueward
trend in $Y-J$ crosses a precipice at the T/Y transition, declining
dramatically ($\approx$0.5~mag) over only half a spectral subclass, from
T9.5 to Y0. This is discussed further in Section~\ref{sec:discussion}.

% - - - - - - - - - - - - - - - - - - - - - - - - - - - - - - - -%
\subsection{Physical Properties from Evolutionary Models \label{sec:properties}}

We use evolutionary models from \citet{2003ApJ...596..587B} and
\citet{2003A&A...402..701B} to derive physical properties for the two
binaries, assuming ages of 1~and 5~Gyr. As inputs, we use both the
$J$-band absolute magnitude and the estimated \Lbol. The uncertainties
in these quantities are propagated into the calculations in a Monte
Carlo fashion.
We use $J$-band as it represents the peak flux of the near-IR spectral
energy distribution. When using the $J$-band absolute magnitudes for our
calculations, we are implicitly using the bolometric corrections from
the underlying model atmospheres.

To compute \Lbol, we adopt a $J$-band bolometric correction of
$1.75\pm0.27$~mag for \wiseoneA\ (T9), \wiseoneB\ (Y0), and \wisetwoB\
($\approx$T9.5), based on the coolest T~dwarfs as described in
\citet{2011ApJ...740..108L}. This is expected to somewhat underestimate
the \Lbol, given the increasingly large fraction of the luminosity
emitted in the mid-IR at colder temperatures
\citep[e.g.][]{2010ApJ...710.1627L}. However, given the lack of actual
bolometric measurements for objects as cool as \wiseoneB, adopting a
fixed value is also conservative.

For the hotter \wisetwoA~(T8), we adopt $J$-band bolometric corrections
from the four T7.5--T8.5 dwarfs with parallaxes: Gl~570D (T7.5; $BC_J =
2.60\pm0.13$~mag), 2MASS~J1217$-$0311 (T7.5, $2.64\pm0.13$~mag), and
2MASS~J0415$-$0935 (T8; $2.54\pm0.13$~mag) from \citet{gol04} and
Wolf~940B (T8.5, $2.09\pm0.12$~mag) from \citet{2010ApJ...720..252L}.
The unweighted average and RMS are $2.47\pm0.26$~mag.

Tables~\ref{table:wiseone}~and~\ref{table:wisetwo} present the results
derived from the evolutionary models. Altogether, the components of
\wiseoneAB\ are remarkably similar to those of \cfbdsAB, with near-IR
absolute magnitudes that differ by only $\approx$0.2~mag; thus, the
inferred physical properties of these two systems are very similar. Note
that the results from the \citet{2003ApJ...596..587B} models should be
somewhat preferred over the Lyon/COND models of
\citet{2003A&A...402..701B}. The predicted near-IR locus of Burrows
\etal\ is a better match to the location of \cfbdsAB\ (Figure~7 of
\citealp{2011ApJ...740..108L}), though the two sets of models basically
give consistent results for the physical properties. Overall, the
inferred temperature of \wiseoneB\ is exceptionally low, 350--470~K
depending on the choice of theoretical model and system age. Its
inferred mass is 5--20~\Mjup, overlapping the old gas-giant planets
found by radial velocity and transit surveys and the young ones found by
direct imaging \citep{2005A&A...438L..25C, 2008ApJ...689L.153L,
  marois08-hr8799bcd, 2010Sci...329...57L}. As expected based on its
brighter estimated absolute magnitudes, we find \wisetwoB\ is somewhat
hotter (420--540~K) and more massive (9--26~\Mjup) than \wiseoneA.

Both components of \wiseoneAB\ would reside in the planetary-mass regime
and have retained their initial deuterium abundance for ages of
$\lesssim$1~Gyr. In fact, the two components of \wiseoneAB\ may possess
different deuterium abundances, as their estimated masses could straddle
the deuterium-burning limit (11--14~\Mjup;
\citealp{2010arXiv1008.5150S}). For \wisetwoAB, the primary likely lies
above the D-burning limit, and the secondary would lie below it for ages
of $\lesssim$1~Gyr. Measuring the differential deuterium abundance in
these binaries would help constrain their ages, analogous to the binary
lithium test proposed by \citet{2005astro.ph..8082L} --- see
\citet{2011ApJ...740..108L} for a discussion of the very similar case of
\cfbdsAB.

% - - - - - - - - - - - - - - - - - - - - - - - - - - - - - - - -%
\subsection{Orbital Periods \label{sec:orbit}}

To estimate the orbital periods of the two new \WISE\
binaries, we adopt the statistical conversion factor between projected
separation and true semi-major axis from \citet{dupuy2011-eccentricity}.
They offer several choices, based on the underlying eccentricity
distribution and degree of completeness to finding binaries by imaging
(``discovery bias'').
We adopt the eccentricity distribution from their compilation of
ultracool visual binaries with high quality orbits and assume no
discovery bias, appropriate for such well-resolved systems.
This gives a multiplicative correction factor of 1.15$^{+0.81}_{-0.31}$
(68.3\% confidence limits) for converting the projected separation into
semi-major axis (compared to $1.10^{+0.92}_{-0.35}$ for a uniform
eccentricity distribution with no discovery bias).
For ages of 1--5~Gyr, the resulting orbital period estimates range from
120--210~yr and 260--430~yr for the two binaries, albeit with
significant uncertainties (Tables~\ref{table:wiseone}
and~\ref{table:wisetwo}). Dynamical masses can be realized from orbit
monitoring covering $\gtrsim$30\% of the period
\citep[e.g.][]{2004A&A...423..341B, liu08-2m1534orbit,
  2009ApJ...706..328D}. Thus, several decades of monitoring will be
needed to determine the visual orbits.

Orbital motion of \cfbdsAB\ is clearly detected in our new data, taken
two~years after the discovery epoch in \citet{2011ApJ...740..108L}.
Comparing the data from the filters with the best astrometry in the two
epochs, we measure a change in separation of $16\pm3$~mas and in
position angle of $8.5\pm1.6\degs$. Such motion is not surprising, given
the $\approx$30~yr orbital period estimated at the time of discovery.

%----------------------------------------------------------------------%
\section{Discussion \label{sec:discussion}}

Theoretical models predict the near-IR spectra of $\approx$300--500~K
objects to be sensitive to new physical processes
\citep[e.g.][]{2003ApJ...596..587B, 2007ApJ...667..537L}.
\citet{2011ApJ...743...50C} have identified changes in near-IR
low-resolution spectra of Y~dwarfs that they ascribe to \ammonia. This
molecule may also be present in high-resolution near-IR spectra of
\ugps\ \citep{2012arXiv1202.6293S, 2011AJ....142..169B}.
Another anticipated signature is the removal of the very broad wings of
the \ion{K}{1} 0.77~\micron\ doublet that dominate the far-red optical
continuum \citep[e.g.][]{2000ApJ...531..438B, 2002ApJ...568..335M}. Down
to $\approx$500~K, \citet{2012ApJ...748...74L} find that the far-red
colors, as tracked by the $izYJ$ filters, continue to be very red with
only a hint of saturation.

\begin{figure}
\vskip -1in
\centerline{\includegraphics[width=6in,angle=90]{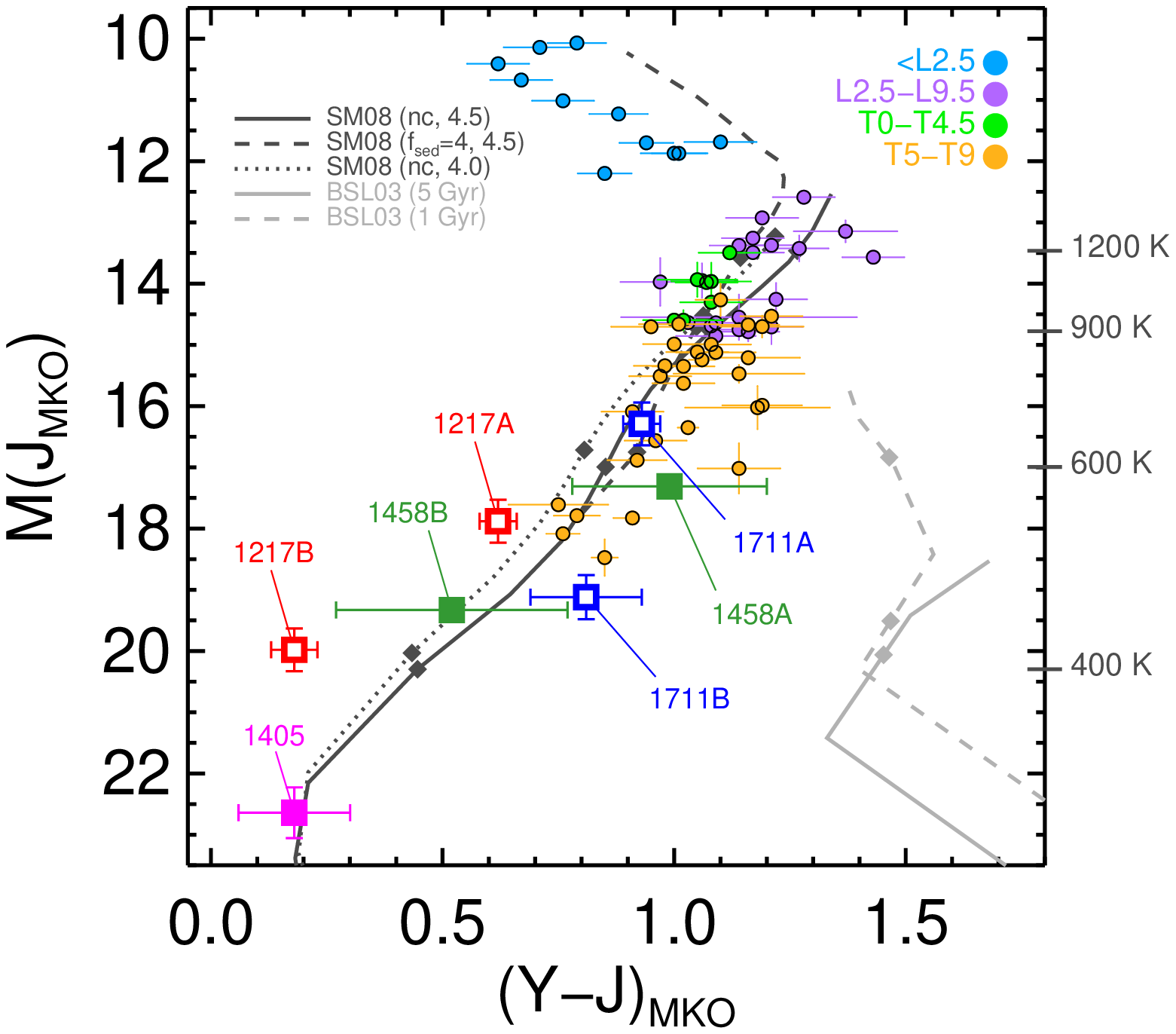}}
%\centerline{\includegraphics[width=6in,angle=90]{plot-cmd.ps}}
\vskip -2ex
\caption{\normalsize Near-IR color-magnitude diagram showing our
  binaries compared to field objects from the compilation of
  \citet{2012arXiv1201.2465D} and to evolutionary models of
  \citet[][light grey]{2003ApJ...596..587B} and \citet[][dark
  grey]{2008ApJ...689.1327S}. For the \WISE\ binaries (open squares),
  the positions are based on the photometric distances to the primary
  components, whereas \cfbds\ and WISE~J1405+5534 (filled squares) have
  parallactic distances from \citet{2012arXiv1201.2465D} and
  \citet{2012arXiv1205.2122K}, respectively. The field sample only shows
  objects with parallax measurements of $S/N > 4$. For the Saumon \&
  Marley models, we plot those that are cloud-free (``nc''; 350--1500~K)
  and that have a modest cloud layer ($f_{sed}=4$; 500--2400~K) for
  $\logg=4.0$ and~4.5~dex. The diamonds on the model tracks demarcate a
  common set of temperatures for all the models, and for the Saumon \&
  Marley \{nc, 4.5\} models (dark grey solid line), the corresponding
  \Teff\ values are listed on the right hand axis. \label{fig:cmd}}
\end{figure}

\begin{figure}
\centerline{\includegraphics[width=6in,angle=90]{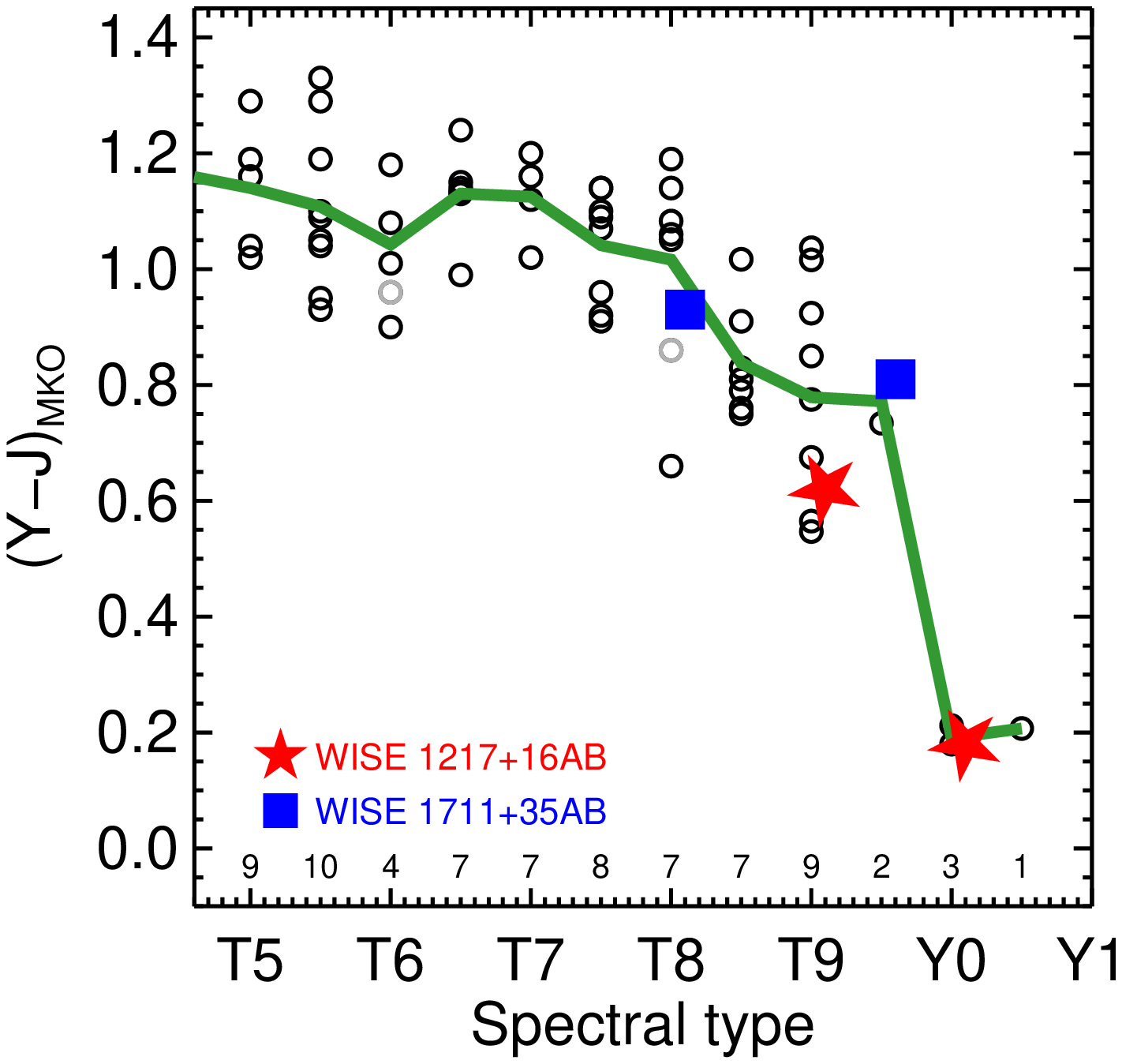}}
%\centerline{\includegraphics[width=6in,angle=90]{plot-yj-colors.ps}}
\vskip -3ex
\caption{$(Y-J)$ color as a function of spectral type for the objects
  plotted in the previous color-color diagram (Figure~\ref{fig:cmd}).
  The median uncertainty for objects with published photometry is
  0.10~mag. Known binaries are plotted in light grey. The components of
  our \WISE\ binaries are slightly displaced horizontally from their
  actual spectra type for clarity. The green line shows the average
  color as a function of spectral type, excluding known binaries. The
  small numbers at the bottom give the sample size used to compute the
  average color of each subclass. (The resolved colors of \cfbdsAB\ are
  not plotted given their larger uncertainties but are consistent with
  objects of similar spectral type. However, as discussed in
  Section~\ref{sec:phot}, the $Y$~and $J$~flux ratios for the system
  also show the large blue color difference between its primary and
  secondary component, like \wiseoneAB\ does.) \label{fig:yj}}
\end{figure}

At somewhat cooler temperatures, below 400~K the Cs, K, Li and Rb
neutral alkalis are expected to be depleted as they turn into chloride
gases and solids, or other halide or hydroxide gases, and Na condenses
as Na$_2$S \citep{1999ApJ...519..793L, 2012ApJ...748...74L}. Given the
strength of the \ion{K}{1} absorption wings, depletion of this element
should be manifested by a change in the broad-band colors.
Figures~\ref{fig:colorcolor},~\ref{fig:cmd},~and~\ref{fig:yj} show that
the $Y-J$ colors of \wiseoneB\ and the other two Y~dwarfs are
significantly bluer than all known T~dwarfs, consistent with the
signature of such depletion. Indeed, the sharp change in $Y-J$ is
reminscent of similar behavior seen in near-IR $JHK$ colors at the L/T
transition due the onset of photospheric methane absorption
\citep[e.g.][]{2004AJ....127.3553K}, though the change in $Y-J$ occurs
over a small range of spectral subclasses.

Based on the photometric distance to its primary, the location of
\wiseoneB\ on the color-magnitude diagram also suggests a temperture low
enough for potassium depletion. Figure~\ref{fig:cmd} compares the
available photometry with updated version of the
\citet{2008ApJ...689.1327S} models from D. Saumon (priv. comm.). These
updated models include the latest \htwo\ collision-induced absorption
and \ammonia\ opacities \citep{2012ApJ...750...74S}. However, the models
are calculated in chemical equilibrium, which overestimates the
strengths of the \ammonia\ absorption in the near-IR (which is
significant) and the mid-IR assuming non-equilibrium conditions are
prevalent, as expected. The cloud-free model sequences show $(Y-J)$
decreasing with decreasing temperature through the T dwarf sequence. The
$J$-band absolute magnitudes of the Y0~dwarfs are comparable to the
$\approx$400~K models, though the observed $Y-J$ colors are somewhat
bluer than the models. The predictions from \citet{2003ApJ...596..587B}
have much redder $Y-J$ colors than the color-magnitude data.

Note that the Saumon \& Marley models include the formation of iron,
silicate and corundum clouds (which are important for L~dwarfs and the
L/T transition) as well as the depletion of the neutral alkalis.
However, they do not include any opacity due to sulphide or chloride
condensates, such as Na$_2$S and KCl which are formed as Na and~K are
depleted. While such condensates have been historically ignored in
ultracool atmosphere modeling, \citet{morley12-clouds} have found the
resulting clouds, especially those of Na$_2$S, significantly affect the
near-IR spectra of mid/late-T~dwarfs and provide a better match to the
$JHK$ data. As shown in Figure~\ref{fig:cmd}, current cloud-free models
agree reasonably well in $\{Y-J, M(J)\}$. But the mismatch for the
latest-type objects may indicate that other factors in addition to
K$\rightarrow$KCl depletion play a role in the $\approx$1~\micron\
fluxes, including non-equilibrium \ammonia\ abundances and the opacity
of sulfate condensates.

%----------------------------------------------------------------------%
\section{Conclusions}

\begin{figure}
%\hskip -0.2in
\includegraphics[height=6.5in,angle=90]{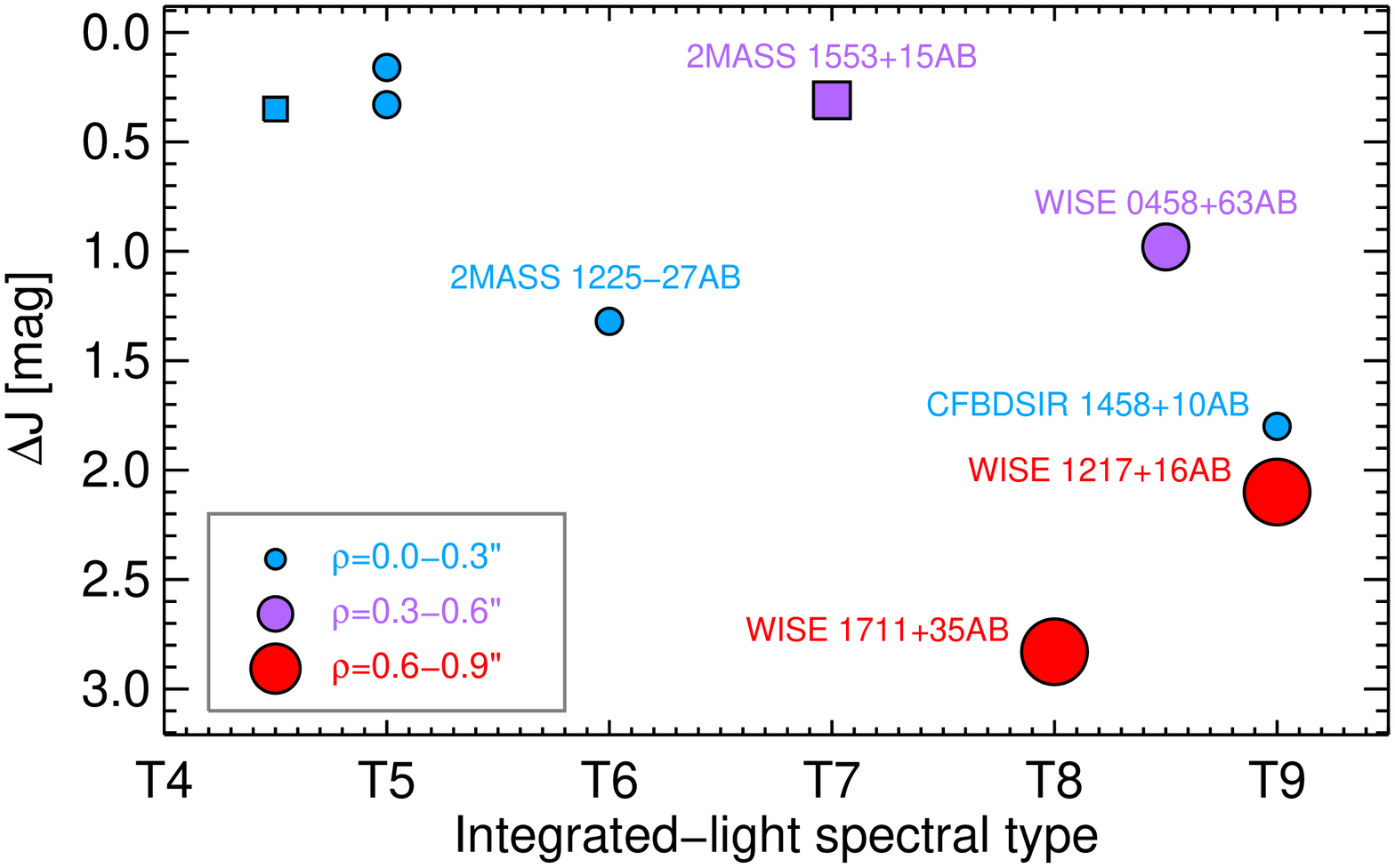}
\caption{\normalsize Integrated-light spectral type versus $J$-band flux
  ratio (a proxy for mass and temperature ratio) for known mid/late-T
  dwarf binaries, based on the compilation of \citet{2010liu-2m1209} and
  more recent discoveries by \citet{2011ApJ...740..108L} and
  \citet{2011arXiv1106.3142G}. (For WISE~J0458+63AB and \cfbdsAB, we use
  the revised spectral types proposed by \citealp{2011ApJ...743...50C}
  of T8.5 and T9, respectively.) The plotting symbol sizes and colors
  represent the projected separation of the binaries. The circles
  represent ground-based photometry, either on the 2MASS or MKO system,
  and the squares represent objects with \HST/NICMOS $F110W$ photometry.
  Only CFBDSIR~1458+10AB \citep{2011ApJ...740..108L} has comparably cold
  components to \wiseoneAB\ and \wisetwoAB, but it has a much tighter
  projected separation (0.11\arcsec).\label{fig:binaries}}
\end{figure}

\wiseoneAB\ and \wisetwoAB\ are remarkably well-resolved binaries
compared to previously known T~dwarf binaries
(Figure~\ref{fig:binaries}), with only the T1+T6 binary \eIndBab\ being
comparably wide in angular separation (0.73\arcsec\ at discovery;
\citealp{2004A&A...413.1029M}). The projected physical separations of 8
and 15~AU for the \WISE\ binaries are also remarkable, with only the
T2.5+T4.0 binary SIMP~J1619275+031350AB being comparable (15~AU
separation; \citealp{2011ApJ...739...48A}) among T~dwarfs. Finally,
their mass ratios of $\approx$0.5 make them very rare among field
ultracool binaries, where nearly equal-mass systems are most prevalent
(\eg, see compilation in \citealp{2010liu-2m1209}). The fact that these
two systems are so unusual in their separation and mass ratios compared
to nearly all previously known substellar binaries opens the question of
whether these initial discoveries among the late-T dwarfs is a harbinger
of a change in binary properties at such cool temperatures. An answer
awaits a larger survey of such targets at comparable sensitivity.

Overall, \wiseoneAB\ appears to be essentially a clone of the coolest
known binary to date, \cfbdsAB, except for the difference in the
projected separation. The blue $Y-J$ color of its secondary component is
unmatched by any previously known T~dwarf, including the primary
component. Synthetic photometry of field Y0--Y0.5~dwarfs
also gives very blue colors, and \cfbdsB\ also shows a much bluer $Y-J$
color than its primary. Examining the complete ensemble of available
$Y-J$ color data, we find a sharp drop to the blue between the T9.5~and
Y0~dwarfs. We suggest this may be the signature of potassium depletion,
which removes a significant opacity source at far-red optical
wavelengths and is expected for objects of $\lesssim$400~K. It is a
fortuitous coincidence that the change in near-IR spectra that
demarcates the T/Y transition (currently suspected to be due to
\ammonia) also coincides with this change in photospheric chemistry.
However, in contrast to the relatively subtle changes in the near-IR
spectra between T9 and Y0, the $Y-J$ color drops by $\approx$0.5~mag.
This suggests that far-red spectra might be more discerning than near-IR
spectra for classification in this temperature regime, as discussed by
\citet{2007ApJ...667..537L} and \citet{2011ApJ...740..108L}.

Future followup will be essential for reaping the full scientific value
of these new binaries. Near-IR low resolution spectra will readily
improve the spectral types. Water clouds are also expected to condense
below $\approx$400~K, further motivating detailed spectroscopic
characterization to search for their signature. The wide separation of
these two binaries means that resolved near-IR and optical spectra can
be obtained in good seeing conditions, without the need for AO.

The binary nature of these systems also provides a unique avenue for
analysis. Among the Y~dwarfs, model atmosphere fitting
\citep{2011ApJ...743...50C} indicates a significant range in
temperatures (350--500~K), surface gravities ($>$1~dex range), and
masses (3--30~\Mjup) that is in contrast to the modest differences in
their near-IR spectra. It is unclear whether such dispersion represents
intrinsic variations in the physical properties of Y~dwarfs, failings in
the model atmospheres, or both. For binaries, the two components have a
common (albeit unknown) metallicity and very similar surface gravity
(Tables~\ref{table:wiseone} and~\ref{table:wisetwo}). Thus resolved
spectra will cleanly probe the effect of decreasing temperature, without
the uncertainties arising from second parameter variations (age/gravity
or metallicity) present in the free-floating population. In a similar
vein, spectrophotometric monitoring of the two components may shed light
on the nature of the clouds at the T/Y transition.

Finally, the photometric distances of 10--20~pc place \wiseoneAB\ and
\wisetwoAB\ easily within reach of near-IR parallax measurements, and we
have begun such monitoring using the Canada-France-Hawaii Telescope
\citep{2012arXiv1201.2465D}. A direct distance will place the four
components in context with other late-T and Y~dwarfs, through both the
absolute magnitude-spectral type relation and color-magnitude diagrams.
Given the unusually small mass ratios of $\approx$0.5 for these systems,
placing the components on the H-R diagram will also uniquely test the
joint accuracy of current evolutionary and atmospheric models, using the
constraint that models must indicate the two components of each binary
are coeval \citep{2010liu-2m1209}.

%----------------------------------------------------------------------%
\acknowledgments

We thank Scott Dahm, Terry Stickel, Jim Lyke, and the Keck Observatory
staff for assistance with LGS AO observing; the Gemini Observatory
staff for obtaining the NIRI photometry through queue observing; and
Kimberly Aller for assistance with the IRTF/SpeX observing.
We thank Andrew Stephens for providing the Gemini/NIRI cold filter
profile; James R. A. Davenport for distributing his IDL implementation
of the \citet{2011BASI...39..289G} ``cubehelix'' color scheme; Caroline
Morley, Jonathan Fortney and collaborators for sharing results in
advance of publication; and Didier Saumon and Isabelle Baraffe for
providing expanded sets of evolutionary models.
This publication makes use of data products from the Wide-Field
Infrared Survey Explorer, which is a joint project of the University
of California, Los Angeles, and the Jet Propulsion
Laboratory/California Institute of Technology, funded by the National
Aeronautics and Space Administration.
Our research has employed the 2MASS data products; NASA's Astrophysical
Data System; the SIMBAD database operated at CDS, Strasbourg, France;
and the Spex Prism Spectral Libraries maintained by Adam Burgasser at
{\tt http://pono.ucsd.edu/$\sim$adam/browndwarfs/spexprism}.
This research was supported by NSF grant AST09-09222 awarded to MCL.
Support for this work to TJD was provided by NASA through
  Hubble Fellowship grant \hbox{{\#}HF-51271.01-A} awarded by the Space
  Telescope Science Institute, which is operated by the Association of
  Universities for Research in Astronomy, Inc., for NASA, under contract
  NAS 5-26555.
Finally, the authors wish to recognize and acknowledge the very
significant cultural role and reverence that the summit of Mauna Kea has
always had within the indigenous Hawaiian community. We are most
fortunate to have the opportunity to conduct observations from this
mountain.

{\it Facilities:} \facility{Keck-2 (LGS/NIRC2), Gemini-North (NIRI),
  IRTF (SpeX)}

%----------------------------------------------------------------------%

%% appendix material should be preceded with a single \appendix command.
%% there should be a \section command for each appendix. mark appendix

%% each appendix (indicated with \section) will be lettered a, b, c, etc.
%% the equation counter will reset when it encounters the \appendix
%% command and will number appendix equations (a1), (a2), etc.

\appendix

\section{$Y$-Band Filter Offsets \label{sec:yband}}

The near-IR spectra of T~dwarfs are highly structured and include strong
absorption bands due to \htwoo. These bands are proximate to the strong
telluric absorption features in the Earth's atmosphere. The telluric
features have served to define the commonly used near-IR photometric
bandpasses, though the exact wavelength ranges of the filters vary.
\citet{2004PASP..116....9S} have computed photometric shifts between
different near-IR filter systems in the ubiquitous $JHK$ bands. The
shifts can be quite large ($\approx$0.4~mag) depending on the particular
filters being considered. The $Y$-band in the $\approx$1~\micron\
atmospheric window is also subject to such effects.

\begin{figure}
\centerline{\includegraphics[width=5in,angle=90]{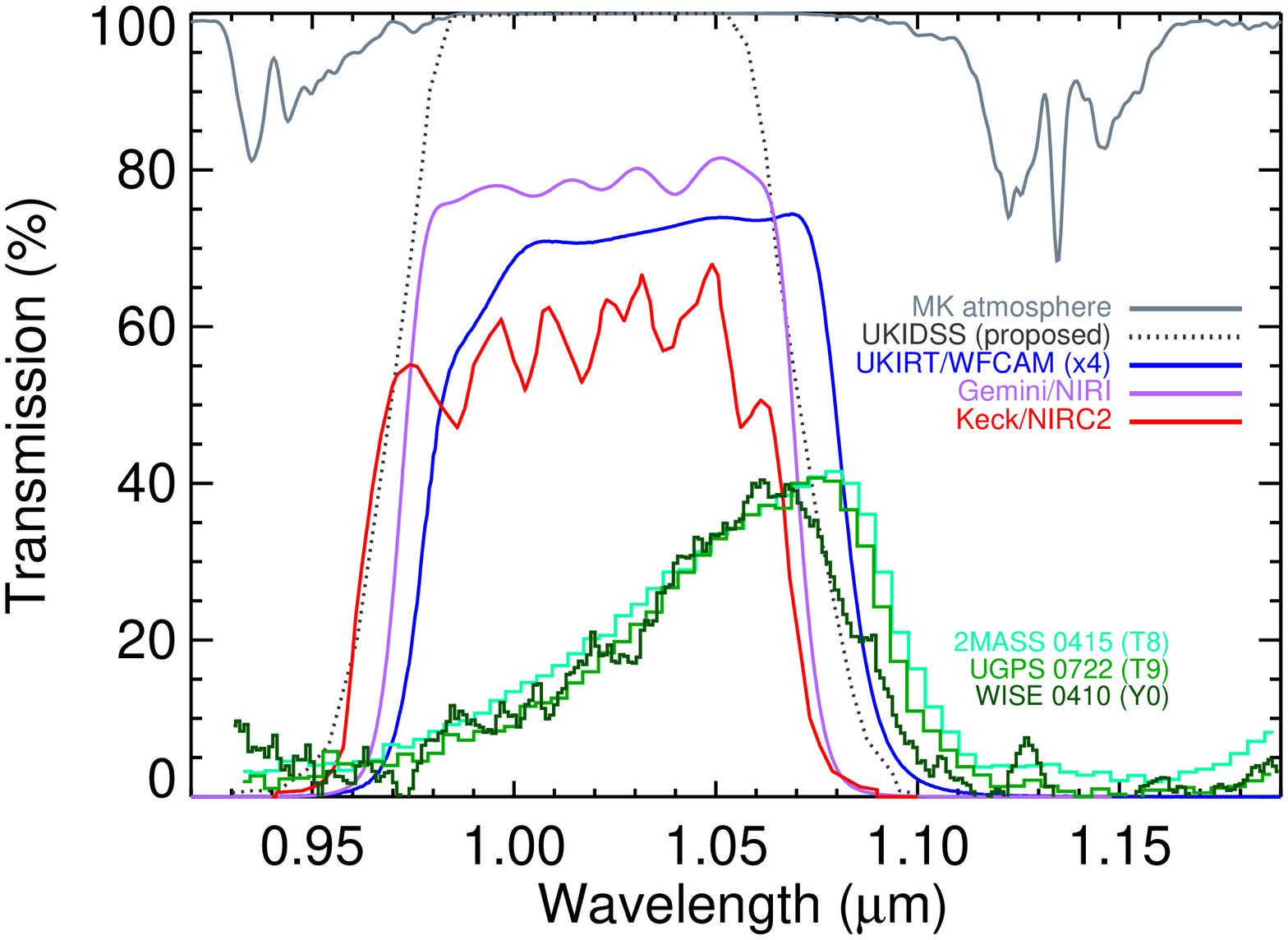}}
%\centerline{\includegraphics[width=5in,angle=90]{yband-with-spectra.ps}}
\vskip -2ex
\caption{\normalsize {Comparison of the $Y$-band filters
    used in this work, from UKIRT/WFCAM, Gemini/NIRI, and Keck/NIRC2.
    The UKIRT/WFCAM curve represents the full combined transmission of
    the atmosphere, telescope, filter, and detector
    \citep{2006MNRAS.367..454H} and has been multiplied by a factor of~4
    to compare with the other two filters. The Gemini/NIRI and
    Keck/NIRC2 curves are the throughput of the filters only and were
    determined for 65~K and 77~K, respectively. We also show an
    atmospheric transmission model for the summit of Mauna Kea (from
    {\tt
      http://www.jach.hawaii.edu/UKIRT/astronomy/utils/atmos-index.html})
    and the original $Y$~band proposed for UKIDSS ({\tt
      http://www.ukidss.org}) based on \citet{2002ASPC..283..369W}. The
    lower curves in green show the T8 and T9~spectroscopic standards and
    the Y0~dwarf WISE~J0410+1502. (The Y0~standard WISE~J1738+2732 of
    \citealp{2011ApJ...743...50C} does not have $Y$-band spectra). The
    spectrum of WISE~J0410+1502 has been smoothed with a boxcar filter
    to boost the S/N, and all three spectra are not plotted below
    0.93~\micron, where their S/N is very low.}
  \label{fig:yband}}
\end{figure}

The measurements in this paper involve $Y$-band photometry from three
instruments: Keck/NIRC2 (resolved photometry of three binaries),
Gemini/NIRI (integrated-light photometry of \wiseoneAB), and UKIRT/WFCAM
(integrated-light photometry of late-T dwarfs from the literature). The
three filters are very similiar, but they are distinct
(Figure~\ref{fig:yband}). The differences are especially significant for
the late-T and Y~dwarfs as the peak of their continuum in this bandpass
falls near the red edge of the filter, so for a given object the
$Y$-band magnitudes will be brightest for the UKIRT/WFCAM filter and
faintest for the Keck/NIRC2 filter. To assess the offsets between
filters, we synthesize colors in the various $Y$-band filters using
spectra of T8--Y0 dwarfs with sufficient blue wavelength coverage,
primarily from \citet{2011arXiv1108.4677K} and
\citet{2011ApJ...743...50C}.

We compute a Gemini-UKIRT shift of $Y_{NIRI} - Y_{WFCAM} =
0.17\pm0.03$~mag, where the uncertainty is the RMS of the synthesized
colors. We apply this small offset to our Gemini/NIRI photometry for a
final value of $Y_{MKO} = 18.38\pm0.04$~mag for \wiseoneAB\
(Table~\ref{table:wiseone}). We assume here that UKIRT/WFCAM serves as
the de facto definition for the $Y$~band of the MKO photometric system,
given that the UKIDSS project carried out with this instrument has
produced the largest set of $Y$-band photometry by far. Note that applying
this shift does not affect one of our main results, namely the very blue
$(Y-J)$ color of \wiseoneB\ compared to previously known objects, since
the secondary is unusually blue relative to both the field objects and
its T9 primary even if we apply no shift at all.

We also compute a Keck-UKIRT shift of $Y_{NIRC2} - Y_{WFCAM} =
0.27\pm0.03$~mag. Figure~\ref{fig:yband} indicates that for Y0~dwarfs,
the WFCAM flux should be systematically different than for the Keck or
Gemini fluxes given the stronger water absorption cutoff in the red edge
of the bandpass for later-type objects, but no convincing offset was
detected between the two Y dwarfs (WISE~J0410+1502 and
WISE~J1541$-$2250) and the T8--T9.5~dwarfs in our compilation,
especially given the low S/N spectra of the latest-type objects.
Therefore, we assume the color offset has no spectral type dependence,
and hence the flux ratios between the binary components in the
$Y_{NIRC2}$ and $Y_{WFCAM}$ filters are assumed to be the same. If there
is such a dependence, Figure~\ref{fig:yband} indicates that shifting
from Keck/NIRC2 to UKIRT/WFCAM would make \wiseoneB\ and \cfbdsB\ even
bluer relative to their primaries.

\begin{figure}
%\hskip -0.2in
\centerline{\includegraphics[height=6.5in,angle=0]{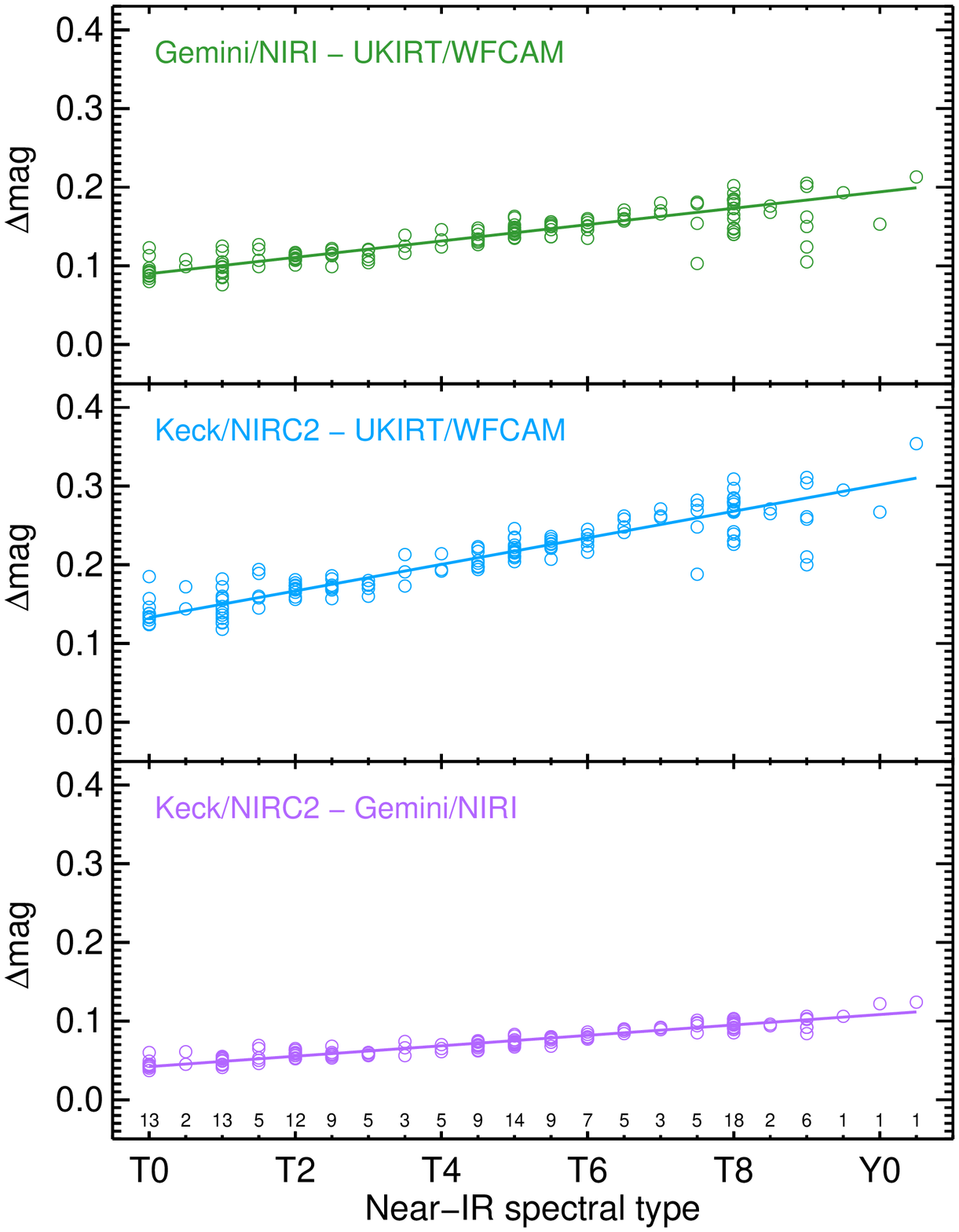}}
%\centerline{\includegraphics[height=6.5in,angle=0]{plot-yband-offsets.ps}}
\vskip 1ex
\caption{\normalsize Synthesized offset between different $Y$-band
  filters for photometry of T~and Y~dwarfs. The label for each subplot
  gives the two relevant filters, \eg, the top panel shows $Y_{NIRI} -
  Y_{WFCAM}$. Note that the S/N of the spectra for the T8--Y0.5 dwarfs
  is quite mixed, leading to scatter in the computed colors. The small
  row of numbers at the bottom gives the sample size for each
  subclass.\label{fig:yoffset}}
\end{figure}

Finally, we synthesize the offsets between these $Y$-band filters over
the entire T~spectral class by adding the low-resolution spectra from
the SpeX Prism Library. Figure~\ref{fig:yoffset} shows the results.
Robust linear fitting to the synthetic photometry gives the following
relations:
$$Y_{NIRI}  - Y_{WFCAM} = 0.090 + 0.010 \times (Tsubclass)   \qquad\sigma = 0.015~{\rm mag}$$
$$Y_{NIRC2} - Y_{WFCAM} = 0.133 + 0.017 \times (Tsubclass)   \qquad\sigma = 0.019~{\rm mag}$$
$$Y_{NIRC2} - Y_{NIRI}  = 0.042 + 0.007 \times (Tsubclass)   \qquad\sigma = 0.006~{\rm mag}$$
where $Tsubclass$ is 0 for T0, 1 for T1, 10 for Y0, etc.\ and $\sigma$
gives the RMS about the fitted relation. The fit is valid for T0 to
Y0.5.

%%======================================================================%%

%% the reference list follows the main body and any appendices.
%% use latex's thebibliography environment to mark up your reference list.
%% note \begin{thebibliography} is followed by an empty set of
%% curly braces.  if you forget this, latex will generate the error
%% "perhaps a missing \item?".
%% cross-referencing. each reference is preceded by a
%% \bibitem command that defines in curly braces the key that corresponds
%% to the key in the \cite commands (see the first section above).
%% make sure that you provide a unique key for every \bibitem or else the
%% paper will not latex. the square brackets should contain
%% the citation text that latex will insert in

%% we have used macros to produce journal name abbreviations.
%% aastex provides a number of these for the more frequently-cited journals.
%% see the author guide for a list of them.

%% note that the style of the \bibitem labels (in []) is slightly
%% different from previous examples.  the natbib system solves a host
%% see the natbib documentation for more details and options.

%\begin{thebibliography}{}
%\end{thebibliography}

%\clearpage
%\bibliography{/users/mliu/tex/bibtex/mliu}
%\bibliographystyle{apj}

%======================================================================%
\clearpage
%% use the figure environment and \plotone or \plottwo to include 

\clearpage

%----------------------------------------------------------------------------------------------------%
\clearpage

\begin{deluxetable}{lcccccccc}
\tablecaption{Keck LGS AO Observations \label{table:keck}}
\rotate
\tabletypesize{\footnotesize}
%\tabletypesize{\small}
\tablewidth{0pt}
\tablecolumns{9}
\tablehead{
  \colhead{Date} &
  \colhead{Filter\tablenotemark{a}} &
  \colhead{$N_{images} \times T_{int}$} &
  \colhead{Airmass} &
  \colhead{FWHM} &
  \colhead{Strehl ratio} &
  \colhead{Separation} &
  \colhead{Position angle} &
  \colhead{$\Delta$mag} \\
  \colhead{(UT)} &
  \colhead{} &
  \colhead{(s)} &
  \colhead{} &
  \colhead{(mas)} &
  \colhead{} &
  \colhead{(mas)} &
  \colhead{(deg)} &
  \colhead{}
}

\startdata
% Date          Filter           Nimgs x Tint      Airmass      FWHM              Strehl              Separation                    PA                        d(mag)

\cutinhead{\wiseoneAB}
%\cline{1-9}    
%%\multicolumn{9}{c}{} \\
%\multicolumn{9}{c}{\bf \wiseoneAB} \\
%%\multicolumn{9}{c}{} \\
%\cline{1-9}
2012-Jan-29   & $Y_{NIRC2}$  &  6 $\times$ 180.0  &  1.03  &  $130\pm5$   &  $0.042\pm0.003$   &  758.2 $\pm$ 1.4 (1.2)  &  14.50 $\pm$ 0.13 (0.10)  &   1.666 $\pm$ 0.011  \\
              & $z_{1.1}$    &  6 $\times$ 300.0  &  1.12  &  $144\pm9$   &  $0.030\pm0.006$   &  759.2 $\pm$ 3.3 (3.2)  &  14.23 $\pm$ 0.26 (0.25)  &   1.69  $\pm$ 0.05   \\
              & $J$         &  7 $\times$ 60.0   &  1.01  &  $115\pm6$   &  $0.069\pm0.007$   &  757.1 $\pm$ 1.6 (1.5)  &  14.32 $\pm$ 0.10 (0.03)  &   2.10  $\pm$ 0.02   \\
              & $CH4{s}$    &  7 $\times$ 60.0   &  1.00  &  $114\pm6$   &  $0.087\pm0.012$   &  758.5 $\pm$ 1.9 (1.7)  &  14.43 $\pm$ 0.22 (0.20)  &   2.15  $\pm$ 0.04   \\
              & $H$         &  7 $\times$ 60.0   &  1.01  &  $119\pm10$  &  $0.082\pm0.017$   &  757.3 $\pm$ 2.5 (2.4)  &  14.37 $\pm$ 0.29 (0.27)  &   2.20  $\pm$ 0.04   \\
              & $K$         &  6 $\times$ 180.0  &  1.06  &  $109\pm3$   &  $0.17\pm0.04$     &  760.7 $\pm$ 6.2 (6.2)  &  14.27 $\pm$ 0.26 (0.24)  &   2.16  $\pm$ 0.13   \\

\cutinhead{\wisetwoAB}
2012-Apr-12   & $Y_{NIRC2}$  &  8 $\times$ 180.0  &  1.05  &   $74\pm7$   &   $0.025\pm0.006$   &  777.2 $\pm$ 8.1 (8.0)  &  328.74 $\pm$ 0.20 (0.19) &   2.71 $\pm$ 0.11    \\
              & $J$         &  7 $\times$ 180.0  &  1.04  &   $61\pm6$   &   $0.059\pm0.004$   &  779.6 $\pm$ 1.1 (1.0)  &  328.40 $\pm$ 0.08 (0.07) &   2.83 $\pm$ 0.06    \\
              & $CH4{s}$    &  8 $\times$ 120.0  &  1.04  &   $55\pm2$   &   $0.137\pm0.018$   &  779.8 $\pm$ 2.7 (2.6)  &  328.51 $\pm$ 0.19 (0.19) &   2.63 $\pm$ 0.11    \\
              & $H$         &  8 $\times$ 120.0  &  1.04  &   $56\pm5$   &   $0.125\pm0.006$   &  780.0 $\pm$ 2.0 (1.9)  &  328.43 $\pm$ 0.23 (0.23) &   2.83 $\pm$ 0.09    \\
2012-Apr-13   & $K$         &  5 $\times$ 180.0  &  1.04  & $60.0\pm1.5$ &   $0.387\pm0.107$   &  780.7 $\pm$ 2.5 (2.4)  &  328.49 $\pm$ 0.09 (0.09) &   3.08 $\pm$ 0.16    \\

\cutinhead{\cfbdsAB}
2012-Apr-13  &  $Y_{NIRC2}$  & $5\times180.0$     &  1.07  &  $65\pm6$    &  $0.057\pm0.040$    &  $125.6\pm1.8$ (1.8)  &  $317.4\pm1.3$(1.3)  &  $1.55\pm0.14$ \\
             &  $J$         & $3\times180.0$     &  1.03  &  $73\pm2$    &  $0.061\pm0.008$    &  $127.2\pm1.4$ (1.4)  &  $318.1\pm1.1$(1.1)  &  $2.02\pm0.07$ \\
             &  $CH4s$      & $7\times180.0$     &  1.12  &  $70\pm6$    &  $0.178\pm0.059$    &  $123.3\pm2.2$ (2.2)  &  $317.8\pm0.8$(0.8)  &  $1.86\pm0.13$ \\

\enddata

\tablecomments{$JHK$ photometry is on the MKO system,
    while the $Y$-band filter is very similar to the UKIDSS filter but
    not the same. (See Appendix for details.) The uncertainties on the
  imaging performance (FWHM and Strehl ratio) are the RMS of the
  measurements from the individual images. For the separation and PA,
  the numbers in parenthesis are the RMS of the measurements before
  adding in the NIRC2 pixel scale and orientation uncertainties (\ie,
  the values relevant for computing the consistency of the astrometry
  between different filters). }

\end{deluxetable}

%----------------------------------------------------------------------------------------------------%
%# BINARY-CALC.IDL:  mliu@sdhcp94.IfA.Hawaii.Edu Wed Apr 11 14:52:03 2012
%----------------------------------------------------------------------------------------------------%
\clearpage
\begin{deluxetable}{lcc}
\vskip -0.5in
\tablecaption{Properties of \wiseoneAB\label{table:wiseone}}
%\tabletypesize{\tiny}
\tabletypesize{\scriptsize}
\tablewidth{0pt}
\tablehead{
\colhead{Property} &
\colhead{Component {A}} &
\colhead{Component {B}}
}

\startdata

Photometric distance (pc)                                     &     \multicolumn{2}{c}{10.5 $\pm$ 1.7}       \\
Projected separation (AU)                                     &     \multicolumn{2}{c}{8.0 $\pm$ 1.3}        \\
 % - - - - - - - - - - - - - - - - %                          &                         &                    \\
$Y_{MKO}$ (mag)                                                &     \multicolumn{2}{c}{18.38 $\pm$ 0.04}                  \\
$J_{MKO}$ (mag)                                                &     \multicolumn{2}{c}{17.83 $\pm$ 0.02\tablenotemark{a}} \\
$H_{MKO}$ (mag)                                                &     \multicolumn{2}{c}{18.18 $\pm$ 0.05\tablenotemark{a}} \\
$K_{MKO}$ (mag)                                                &     \multicolumn{2}{c}{18.80 $\pm$ 0.04}                  \\
   % - - - - - - - - - - - - - - - - %                        &                          &                       \\
Near-IR spectral type                                         &     T9 $\pm$ 0.5         &     Y0 $\pm$ 0.5      \\
   % - - - - - - - - - - - - - - - - %                        &                          &                       \\
$Y_{MKO}$ (mag)                                                &  18.59 $\pm$ 0.04        &  20.26 $\pm$ 0.04     \\    % shifting Gemini/NIRI to UKIRT/WFCAM 
$J_{MKO}$ (mag)                                                &  17.98 $\pm$ 0.02        &  20.08 $\pm$ 0.03     \\
$H_{MKO}$ (mag)                                                &  18.31 $\pm$ 0.05        &  20.51 $\pm$ 0.06     \\
$K_{MKO}$ (mag)                                                &  18.94 $\pm$ 0.04        &  21.10 $\pm$ 0.12     \\
   % - - - - - - - - - - - - - - - - %                        &                          &                       \\
$(Y-J)_{MKO}$ (mag)                                            &  \phs0.62 $\pm$ 0.04     &  \phs0.18 $\pm$ 0.05  \\    % shifting Gemini/NIRI to UKIRT/WFCAM 
$(J-H)_{MKO}$ (mag)                                            &   $-$0.34 $\pm$ 0.05     &   $-$0.44 $\pm$ 0.07  \\
$(H-K)_{MKO}$ (mag)                                            &   $-$0.62 $\pm$ 0.07     &   $-$0.58 $\pm$ 0.14  \\
$(J-K)_{MKO}$ (mag)                                            &   $-$0.96 $\pm$ 0.05     &   $-$1.02 $\pm$ 0.12  \\
   % - - - - - - - - - - - - - - - - %                        &                          &                       \\
$M(Y_{MKO})$ (mag)                                             &  18.49 $\pm$ 0.35        &  20.16 $\pm$ 0.35     \\     % shifting Gemini/NIRI to UKIRT/WFCAM 
$M(J_{MKO})$ (mag)                                             &  17.88 $\pm$ 0.35        &  19.98 $\pm$ 0.35     \\
$M(H_{MKO})$ (mag)                                             &  18.21 $\pm$ 0.35        &  20.41 $\pm$ 0.36     \\
$M(K_{MKO})$ (mag)                                             &  18.84 $\pm$ 0.35        &  21.00 $\pm$ 0.37     \\
   % - - - - - - - - - - - - - - - - %                        &                          &                       \\
$\log(L_{bol}/L_{\odot})$ [from $J$ band]                       &  $-$5.95 $\pm$ 0.18      &  $-$6.79 $\pm$ 0.18   \\
    % - - - - - - - - - - - - - - - - %                        &                          &                       \\
$\log(L_{bol,B} / L_{bol,A})$ [from $J$ band]                    &     \multicolumn{2}{c}{$-$0.84 $\pm$ 0.15}        \\
    % - - - - - - - - - - - - - - - - %                        &                          &                       \\

% - - - - - - - - - - - - - - - - - - - - - - - - - - - - - - - - -%
% WISE 1217 - BSL03: from M(J), using CALC-MASSES.IDL, two different ages
% - - - - - - - - - - - - - - - - - - - - - - - - - - - - - - - - -%
\cutinhead{Evolutionary model results for 1.0 and 5.0~Gyr: Using Burrows \etal\ (2003) models and $M(J)$}
Mass ($M_{Jup}$)                                                 &  11.5 $\pm$ 1.1, 29 $\pm$ 3       &   7.4 $\pm$ 0.5, 18.4 $\pm$ 1.0  \\
$q\ (\equiv M_B/M_A)$                                            &   \multicolumn{2}{c}{0.64 $\pm$ 0.08, 0.63 $\pm$ 0.08}    \\
$T_{eff} (K)$                                                    &   490 $\pm$ 30, 530 $\pm$ 30       &    381 $\pm$ 13,  402 $\pm$ 11  \\  
$\log(g)$ (cgs)                                                 &   4.39 $\pm$ 0.03, 4.95 $\pm$ 0.05 &    4.18 $\pm$ 0.03,  4.68 $\pm$ 0.03  \\
Orbital period (yr)                                             &   \multicolumn{2}{c}{$210^{+370}_{-50}$, $130^{+230}_{-30}$}  \\

% - - - - - - - - - - - - - - - - - - - - - - - - - - - - - - - - -%
% WISE 1217 - Lyon/COND: from J-band absolute magnitude, using CALC-MASSES.IDL
% - - - - - - - - - - - - - - - - - - - - - - - - - - - - - - - - -%
\cutinhead{Evolutionary model results for 1.0, 5.0~Gyr: Using Lyon/COND models and $M(J)$} 
Mass ($M_{Jup}$)                                                &  14.4 $\pm$ 1.8, 35 $\pm$ 3        &   8.3 $\pm$ 0.9, 20 $\pm$ 2  \\
$q\ (\equiv M_B/M_A)$                                          &   \multicolumn{2}{c}{0.58 $\pm$ 0.10, 0.57 $\pm$ 0.08}     \\
$T_{eff} (K)$                                                   &  610 $\pm$ 40, 660 $\pm$ 40        &    430 $\pm$ 30, 470 $\pm$ 30    \\
$\log(g)$ (cgs)                                                &  4.54 $\pm$ 0.07, 5.07 $\pm$ 0.05  &   4.27 $\pm$ 0.05, 4.77 $\pm$ 0.05 \\
Orbital period (yr)                                            &   \multicolumn{2}{c}{$190^{+340}_{-50}$, $120^{+220}_{-30}$}  \\

% - - - - - - - - - - - - - - - - - - - - - - - - - - - - - - - - -%
% WISE 1217 - Lyon/COND: from Lbol, using CALC-MASSES.IDL, 2 ages
% - - - - - - - - - - - - - - - - - - - - - - - - - - - - - - - - -%
\cutinhead{Evolutionary model results for 1.0 and 5.0~Gyr: Using Lyon/COND models and \Lbol} 
Mass ($M_{Jup}$)                                                     &   13 $\pm$ 3, 33 $\pm$ 5      &   5.5 $\pm$ 1.2,  13 $\pm$ 3 \\
$q\ (\equiv M_B/M_A)$                                               &   \multicolumn{2}{c}{0.42 $\pm$ 0.22, 0.40 $\pm$ 0.12}         \\
$T_{eff} (K)$                                                        &   580 $\pm$ 70,  630 $\pm$ 70  &  350 $\pm$ 40,  370 $\pm$ 50 \\
$\log(g)$ (cgs)                                                      &  4.47 $\pm$ 0.10,  5.04 $\pm$ 0.09  & 4.07 $\pm$ 0.10,  4.54 $\pm$ 0.11  \\
Orbital period (yr)                                                  &   \multicolumn{2}{c}{$210^{+380}_{-50}$, $130^{+240}_{-30}$}          \\

\enddata

\tablenotetext{a}{\citet{2011arXiv1108.4677K}}

\end{deluxetable}

%--------------------------------------------------------------------------------%
%# BINARY-CALC.IDL:  mliu@notebook-3.local Tue Jun  5 18:00:00 2012
%--------------------------------------------------------------------------------%
\begin{deluxetable}{lcc}
\tablecaption{Properties of WISE~J1711+3500AB\label{table:wisetwo}}
\tablewidth{0pt}
\tabletypesize{\scriptsize}
\tablehead{
\colhead{Property} &
\colhead{Component {A}} &
\colhead{Component {B}}
}

\startdata

Photometric distance (pc)                                     &     \multicolumn{2}{c}{19 $\pm$ 3}        \\
Projected separation (AU)                                     &     \multicolumn{2}{c}{15 $\pm$ 2}        \\
 % - - - - - - - - - - - - - - - - %                          &                         &                 \\
$Y_{MKO}$ (mag)                                                &     \multicolumn{2}{c}{18.51 $\pm$ 0.03\tablenotemark{a}} \\
$J_{MKO}$ (mag)                                                &     \multicolumn{2}{c}{17.59 $\pm$ 0.03} \\
$H_{MKO}$ (mag)                                                &     \multicolumn{2}{c}{18.05 $\pm$ 0.03\tablenotemark{a}} \\
$K_{MKO}$ (mag)                                                &     \multicolumn{2}{c}{18.24 $\pm$ 0.03\tablenotemark{a}} \\
   % - - - - - - - - - - - - - - - - %                        &                            &                       \\
Near-IR spectral type                                         &    T8 $\pm$ 0.5            &   (T9.5 $\pm$ 0.5)\tablenotemark{b}     \\
   % - - - - - - - - - - - - - - - - %                        &                            &                       \\
$Y_{MKO}$ (mag)                                                &  18.60 $\pm$ 0.03          &  21.31 $\pm$ 0.11     \\
$J_{MKO}$ (mag)                                                &  17.67 $\pm$ 0.03          &  20.50 $\pm$ 0.06     \\
$H_{MKO}$ (mag)                                                &  18.13 $\pm$ 0.03          &  20.96 $\pm$ 0.09     \\
$K_{MKO}$ (mag)                                                &  18.30 $\pm$ 0.03          &  21.38 $\pm$ 0.15     \\
   % - - - - - - - - - - - - - - - - %                        &                            &                       \\
$(Y-J)_{MKO}$ (mag)                                            &  \phs0.93 $\pm$ 0.04       &  \phs0.81 $\pm$ 0.12  \\
$(J-H)_{MKO}$ (mag)                                            &   $-$0.46 $\pm$ 0.04       &   $-$0.46 $\pm$ 0.11  \\
$(H-K)_{MKO}$ (mag)                                            &   $-$0.17 $\pm$ 0.04       &   $-$0.42 $\pm$ 0.18  \\
$(J-K)_{MKO}$ (mag)                                            &   $-$0.63 $\pm$ 0.04       &   $-$0.88 $\pm$ 0.17  \\
   % - - - - - - - - - - - - - - - - %                        &                            &                       \\
$M(Y_{MKO})$ (mag)                                             &  17.22 $\pm$ 0.35          &  19.93 $\pm$ 0.37     \\
$M(J_{MKO})$ (mag)                                             &  16.29 $\pm$ 0.35          &  19.12 $\pm$ 0.36     \\
$M(H_{MKO})$ (mag)                                             &  16.75 $\pm$ 0.35          &  19.58 $\pm$ 0.36     \\
$M(K_{MKO})$ (mag)                                             &  16.92 $\pm$ 0.35          &  20.00 $\pm$ 0.38     \\
   % - - - - - - - - - - - - - - - - %                        &                            &                       \\
$\log(L_{bol}/L_{\odot})$  [from $J$ band]                      &  $-$5.60 $\pm$ 0.17        &  $-$6.45 $\pm$ 0.18   \\
%   % - - - - - - - - - - - - - - - - %                        &                            &                       \\
$\log(L_{bol,B} / L_{bol,A})$  [from $J$ band]                   &      \multicolumn{2}{c}{$-$0.84 $\pm$ 0.15}        \\
%   % - - - - - - - - - - - - - - - - %                        &                            &                       \\

% - - - - - - - - - - - - - - - - - - - - - - - - - - - - - - - - -%
% WISE 1711 - BSL03: from M(J), using CALC-MASSES.IDL, two different ages
% - - - - - - - - - - - - - - - - - - - - - - - - - - - - - - - - -%
\cutinhead{Evolutionary model results for 1.0 and 5.0~Gyr: Using Burrows \etal\ (2003) models and $M(J)$}
Mass ($M_{Jup}$)                                                 &  19 $\pm$ 3, 44 $\pm$ 4       &   8.7 $\pm$ 0.8, 22 $\pm$ 2  \\
$q\ (\equiv M_B/M_A)$                                            &   \multicolumn{2}{c}{ 0.46 $\pm$ 0.09, 0.50 $\pm$ 0.07 }    \\
$T_{eff} (K)$                                                    &   675 $\pm$ 50, 680 $\pm$ 40       &   420 $\pm$ 20, 440 $\pm$ 30   \\  
$\log(g)$ (cgs)                                                 &   4.65 $\pm$ 0.07, 5.18 $\pm$ 0.05 &  4.26 $\pm$ 0.03, 4.76 $\pm$ 0.05 \\
Orbital period (yr)                                             &   \multicolumn{2}{c}{430$^{+770}_{-100}$, 280$^{+500}_{-70}$}  \\

% - - - - - - - - - - - - - - - - - - - - - - - - - - - - - - - - -%
% WISE 1711 - Lyon/COND: from J-band absolute magnitude, using CALC-MASSES.IDL
% - - - - - - - - - - - - - - - - - - - - - - - - - - - - - - - - -%
\cutinhead{Evolutionary model results for 1.0, 5.0~Gyr: Using Lyon/COND models and $M(J)$} 
Mass ($M_{Jup}$)                                                &  23 $\pm$ 2, 48 $\pm$ 3        &  10.7 $\pm$ 0.9, 26 $\pm$ 2  \\
$q\ (\equiv M_B/M_A)$                                          &   \multicolumn{2}{c}{ 0.46 $\pm$ 0.06, 0.54 $\pm$ 0.05 }     \\
$T_{eff} (K)$                                                   &  810 $\pm$ 50, 870 $\pm$ 60        &   500 $\pm$ 30, 540 $\pm$ 30  \\
$\log(g)$ (cgs)                                                &  4.78 $\pm$ 0.05, 5.27 $\pm$ 0.04  &  4.39 $\pm$ 0.04, 4.90 $\pm$ 0.05  \\
Orbital period (yr)                                            &   \multicolumn{2}{c}{390$^{+700}_{-90}$, 260$^{+480}_{-60}$}  \\

% - - - - - - - - - - - - - - - - - - - - - - - - - - - - - - - - -%
% WISE 1711 - Lyon/COND: from Lbol, using CALC-MASSES.IDL, 2 ages
% - - - - - - - - - - - - - - - - - - - - - - - - - - - - - - - - -%
\cutinhead{Evolutionary model results for 1.0 and 5.0~Gyr: Using Lyon/COND models and \Lbol} 
Mass ($M_{Jup}$)                                                     &   19 $\pm$ 3, 44 $\pm$ 6             &  8.1 $\pm$ 1.6, 20 $\pm$ 4 \\
$q\ (\equiv M_B/M_A)$                                               &   \multicolumn{2}{c}{ 0.43 $\pm$ 0.12, 0.45 $\pm$ 0.12  } \\
$T_{eff} (K)$                                                        &   730 $\pm$ 80,  800 $\pm$ 90        &   430 $\pm$ 50, 460 $\pm$ 50 \\
$\log(g)$ (cgs)                                                      &  4.70 $\pm$ 0.09,  5.21 $\pm$ 0.08  &   4.26 $\pm$ 0.10, 4.76 $\pm$ 0.10 \\
Orbital period (yr)                                                  &   \multicolumn{2}{c}{430$^{+780}_{-110}$, 280$^{+510}_{-70}$}          \\

\enddata

\tablenotetext{a}{Synthesized from our $J$-band photometry and IRTF/SpeX
  spectrum (Section~\ref{sec:spex}).}

\tablenotetext{b}{Estimated from $H$-band absolute magnitude
  (Section~\ref{sec:spectra}).}

\end{deluxetable}

%------------------------------------------------------------%
%# BINARY-CALC.IDL:  mliu@sdhcp94.IfA.Hawaii.Edu Fri Jun 15 01:25:41 2012
%------------------------------------------------------------%
\clearpage
\begin{deluxetable}{lcc}
\tablecaption{Resolved Photometry of CFBDSIR J1458+1013AB\label{table:cfbds}}
\tablewidth{0pt}
\tabletypesize{\scriptsize}
\tablehead{
\colhead{Property} &
\colhead{Component {A}} &
\colhead{Component {B}}
}

\startdata

Distance (pc)                                                 &     \multicolumn{2}{c}{31.9 $\pm$ 2.5\tablenotemark{a}}    \\
Projected separation (AU)                                     &     \multicolumn{2}{c}{3.5 $\pm$ 0.3\tablenotemark{b}}        \\
   % - - - - - - - - - - - - - - - - %                        &                            &                       \\
$Y_{MKO}$ (mag)                                                &     \multicolumn{2}{c}{20.58 $\pm$ 0.21\tablenotemark{c}} \\
$J_{MKO}$ (mag)                                                &     \multicolumn{2}{c}{19.67 $\pm$ 0.02\tablenotemark{d}} \\
$H_{MKO}$ (mag)                                                &     \multicolumn{2}{c}{20.06 $\pm$ 0.10\tablenotemark{d}} \\
$K_{MKO}$ (mag)                                                &     \multicolumn{2}{c}{20.50 $\pm$ 0.24\tablenotemark{d}} \\
   % - - - - - - - - - - - - - - - - %                        &                            &                       \\
Near-IR spectral type                                    &  T9 $\pm$ 0.5\tablenotemark{e}  &  (Y0 $\pm$ 0.5)\tablenotemark{f}  \\
   % - - - - - - - - - - - - - - - - %                        &                            &                       \\
$Y_{MKO}$ (mag)                                                &  20.81 $\pm$ 0.21          &  22.36 $\pm$ 0.24     \\
$J_{MKO}$ (mag)                                                &  19.83 $\pm$ 0.02          &  21.85 $\pm$ 0.06     \\
$H_{MKO}$ (mag)                                                &  20.18 $\pm$ 0.10          &  22.51 $\pm$ 0.16     \\
$K_{MKO}$ (mag)                                                &  20.63 $\pm$ 0.24          &  22.83 $\pm$ 0.30     \\
   % - - - - - - - - - - - - - - - - %                        &                            &                       \\
$(Y-J)_{MKO}$ (mag)                                            &  \phs0.99 $\pm$ 0.21       &  \phs0.52 $\pm$ 0.25  \\
$(J-H)_{MKO}$ (mag)                                            &   $-$0.35 $\pm$ 0.10       &   $-$0.66 $\pm$ 0.17  \\
$(H-K)_{MKO}$ (mag)                                            &   $-$0.45 $\pm$ 0.26       &   $-$0.32 $\pm$ 0.34  \\
$(J-K)_{MKO}$ (mag)                                            &   $-$0.81 $\pm$ 0.24       &   $-$0.99 $\pm$ 0.30  \\
%   % - - - - - - - - - - - - - - - - %                        &                            &                       \\
$M(Y_{MKO})$ (mag)                                             &  18.29 $\pm$ 0.27          &  19.84 $\pm$ 0.29     \\
$M(J_{MKO})$ (mag)                                             &  17.31 $\pm$ 0.17          &  19.33 $\pm$ 0.18     \\
$M(H_{MKO})$ (mag)                                             &  17.66 $\pm$ 0.20          &  19.99 $\pm$ 0.23     \\
$M(K_{MKO})$ (mag)                                             &  18.11 $\pm$ 0.30          &  20.31 $\pm$ 0.34     \\
   % - - - - - - - - - - - - - - - - %                        &                            &                       \\
$\log(L_{bol}/L_{\odot})$  [from $J$ band]                      &  $-$5.72 $\pm$ 0.13        &  $-$6.53 $\pm$ 0.13   \\
   % - - - - - - - - - - - - - - - - %                        &                            &                       \\
$\log(L_{bol,B} / L_{bol,A})$  [from $J$ band]                  &   \multicolumn{2}{c}{$-$0.81 $\pm$ 0.15}           \\

\enddata

\tablenotetext{a}{\citet{2012arXiv1201.2465D}}

\tablenotetext{b}{Epoch 2010 July 08 UT}

\tablenotetext{c}{UKIDSS Data Release 8}

\tablenotetext{d}{\citet{2010A&A...518A..39D}}

\tablenotetext{e}{Classification by \citet{2011ApJ...743...50C} based on
  \citet{2011ApJ...740..108L} spectrum.}

\tablenotetext{f}{Estimated from $H$-band absolute magnitude
  (Section~\ref{sec:spectra}).} 

\tablecomments{Resolved $Y$- and $J$-band photometry based on new flux
  ratio measurements in this paper, while the flux ratios in the other
  bandpasses and all the integrated-light photometry come from
  \citet{2011ApJ...740..108L}.}

\end{deluxetable}

%------------------------------------------------------------------------------------------------------------------------%
% UKIDSS Photometry
%------------------------------------------------------------------------------------------------------------------------%
\clearpage

\begin{deluxetable}{lcccccc}
\tablecaption{UKIDSS DR9 Photometry of \WISE\ Brown Dwarfs \label{table:ukidss}}
\tabletypesize{\small}
\tablewidth{0pt}
%\tablecolumns{2}
\tablehead{
  \colhead{Object} &
  \colhead{Ref.} &
  \colhead{SpT} &
  \colhead{$Y_{MKO}$} &
  \colhead{$J_{MKO}$} &
  \colhead{$H_{MKO}$} &
  \colhead{$K_{MKO}$} \\
  \colhead{} &
  \colhead{} &
  \colhead{} &
  \colhead{(mag)} &
  \colhead{(mag)} &
  \colhead{(mag)} &
  \colhead{(mag)}
}

\startdata
  WISE J0049+0441                  &  1   &  L9    &  16.903 $\pm$ 0.012  &  15.767 $\pm$ 0.007  &  14.801 $\pm$ 0.006  &  14.131 $\pm$ 0.005  \\  
  WISE J0254+0223\tablenotemark{a} &  2,3 &  T8    &  16.999 $\pm$ 0.014  &  15.916 $\pm$ 0.008  &  16.29  $\pm$ 0.02   &  16.73  $\pm$ 0.05   \\      % the Screamer
  WISE J0750+2725                  &  1   &  T8.5  &  19.75  $\pm$ 0.09   &  18.73  $\pm$ 0.05   &  19.00  $\pm$ 0.06\tablenotemark{b}   &         \ldots       \\      % H-band from Kirkpatrick11
  WISE J0929+0409                  &  1   &  T6.5  &  18.00  $\pm$ 0.02   &  16.868 $\pm$ 0.014  &  17.37  $\pm$ 0.07   &  17.40 $\pm$ 0.09  \\      
  WISE J1311+0122                  &  1   &  T9    &  19.89  $\pm$ 0.10   &  18.97  $\pm$ 0.08   &        \ldots        &         \ldots       \\
  WISE J2226+0440                  &  1   &  T8    &  18.04  $\pm$ 0.03   &  16.899 $\pm$ 0.019  &  17.45  $\pm$ 0.07   &  17.24 $\pm$ 0.09  \\  
  WISE J2344+1034                  &  1   &  T9    &  19.88  $\pm$ 0.12   &  18.84  $\pm$ 0.09   &  19.24  $\pm$ 0.29   &         \ldots       \\
\enddata

\tablenotetext{a}{a.k.a.\ PSO~J043.5+02}

\tablenotetext{b}{$H$-band photometry from \citet{2011arXiv1108.4677K}.}

\tablerefs{(1)~\citet{2011arXiv1108.4677K},
  (2)~\citet{2011ApJ...740L..32L}, (3)~\citet{2011A&A...532L...5S}.}

\end{deluxetable}

\end{document}